\documentclass[trackchanges,preprint2]{aastex7}
\usepackage{amsmath}
\usepackage{soul}
\usepackage{longtable} % Uzun tablolar için gerekli paket
\usepackage{caption}   % Daha gelişmiş başlık kontrolü
\usepackage{subcaption}
\usepackage{graphicx}
\usepackage{float}
\usepackage{booktabs} % daha şık çizgiler için (opsiyonel)

\begin{document}

\title{Classification of X-ray Binaries in NGC 6946 with the use of Multiwavelength Data}

\author[orcid=0000-0001-8986-8022,sname='Avdan']{Senay Avdan}
\affiliation{Sabanc{\i} University, Faculty of Engineering and Natural Science, 34956, Istanbul, T\"{u}rkiye}
\affiliation{T\"{u}rkiye National Observatories, DAG, 25050, Erzurum, T\"{u}rkiye}
\email[show]{kayaci.s@gmail.com}  

\author[orcid=0000-0001-8694-1503,sname='Avdan']{Hasan Avdan} 
\affiliation{T\"{u}rkiye National Observatories, DAG, 25050, Erzurum, T\"{u}rkiye}
\email{avdan.hsn@gmail.com}

\author[0000-0002-6909-192X,sname=Sonbas]{Eda Sonbas}
\affiliation{Department of Physics, The George Washington University, Washington, DC 20052, USA}
%\affiliation{Sabanc{\i} University, Faculty of Engineering and Natural Science, 34956, Istanbul, T\"{u}rkiye}
\email{edasonbas@gmail.com}

\author[0000-0002-5422-4873,sname=Dhuga]{Kalvir S. Dhuga}
\affiliation{Department of Physics, The George Washington University, Washington, DC 20052, USA}
\email{dhuga@gwu.edu}

\begin{abstract}

We identified X-ray binary (XRB) candidates in the galaxy NGC 6946 using archival {\it Chandra} data. A total of 98 XRB candidates were detected within the 0.3$-$8 keV energy range, with luminosities between 10$^{36}$ and 10$^{40}$ erg~s$^{-1}$. X-ray color-color diagrams were constructed for these sources, classifying 57 as high-mass X-ray binaries (HMXBs) and 8 as low-mass X-ray binaries (LMXBs). The X-ray temporal variability of the sources was analyzed, revealing that 40 sources exhibit variability and 26 are classified as transient sources. Possible optical counterparts for 57 sources were identified using {\it Hubble Space Telescope} ({\it HST}) images. Based on the color-magnitude diagrams (CMDs), 40 sources were found to have masses greater than 8M$_{\odot}$ and were classified as HMXBs. Seventeen sources with masses between 5 and 8M$_{\odot}$ were classified as intermediate-mass X-ray binaries (IMXBs). The remaining sources, for which optical counterparts could not be identified, were assumed to be LMXBs. The X-ray luminosity functions (XLFs) derived for HMXBs and LMXBs, identified through both X-ray and optical classification, were best described by a single power-law model ($\alpha \sim$ 1.3$-$1.6). We supplemented traditional classification methods with a machine learning (ML) approach based on multiwavelength photometry. While traditional methods identified no AGN candidates, the ML  approach detected a significant AGN population. The spatial resolution of the observing instruments and the limits of the training database hinder the reliability of the ML classifications, thus highlighting a key challenge in combining these methods for extragalactic X-ray source analysis.

\end{abstract}



\section{Introduction}

A substantial fraction of X-ray emission from galaxies is thought to originate from X-ray binaries (XRBs). These systems consist of a compact object, either a neutron star or a black hole, with a stellar companion (the donor star) \citep{2002ApJS..142..239Z, 2020MNRAS.492.5075A}. The X-ray emission occurs as material accretes onto the compact object from the companion star. XRBs are classified into two main types based on the mass of the companion star (M$_{d}$) and mass transfer processes: high-mass X-ray binaries (HMXBs) M$_{d}$ $\geq$ 8 M$_{\odot}$ and low-mass X-ray binaries (LMXBs)  with M$_{d}$ $\leq$ 1 M$_{\odot}$. HMXBs typically feature a young, massive companion star that transfers material on to the compact object through stellar winds, whereas LMXBs consist of an older, low-mass companion star transferring material via Roche-lobe overflow. These sources are major contributors to the luminosity of X-rays in a galaxy and serve as crucial tools for studying stellar evolution, compact objects, and binary interactions \citep{1997xrb..book.....L, 2003MNRAS.339..793G, 2006ARA&A..44..323F, 2017hsn..book.1499C, 2021NewA...8501514I, 2021ApJ...922..178R}. 
\\
\\
In addition, there is another group \citep{2000ApJ...530L..93T, 2002ApJ...565.1107P}: Intermediate-mass X-ray binaries (IMXBs), which represent a transitional class of XRBs whose companion stars have masses between those typically found in LMXBs and HMXBs, ranging from about 2 to 8 M$_{\odot}$. In IMXBs, the mass transfer mechanism can vary depending on the evolutionary state of the donor star and the orbital configuration of the system. In some cases, the companion may overflow its Roche lobe, while in others stellar winds may dominate. IMXBs may exhibit similarities to HMXBs or LMXBs depending on the mass and evolutionary stage of the donor star. These systems are rarer and less studied than LMXBs and HMXBs, but they provide critical insights into binary evolution pathways and the role of intermediate-mass stars in shaping X-ray emission \citep{2003ApJ...597.1036P,2020A&A...642A.174M,  2021ApJ...912...31H, 2023ApJ...945....7G}.
\\
\\
HMXBs and LMXBs differ significantly in their physical characteristics, including their mechanisms of matter transfer, and they are associated with distinct galactic environments. HMXBs, with relatively short lifetimes ($\leq$ 10$^{7}$ yr) determined by the mass of their companion stars, are predominantly found in star-forming regions \citep{2004MNRAS.349..146G, 2006ARA&A..44..323F, 2006csxs.book..623T, 2012MNRAS.419.2095M, 2013A&A...553A...7D}. LMXBs have much longer lifetimes ($\geq$ 10$^{9}$ yr), are linked to the total stellar mass of a galaxy, and are mostly found in the galactic bulge and globular clusters. As a result, LMXBs offer insights into the older stellar populations of a galaxy and its early evolution \citep{2004MNRAS.349..146G, 2010ApJ...724..559L, 2012A&A...546A..36Z}. The classification and study of XRBs is crucial in probing the various evolutionary stages of a galaxy \citep{2004MNRAS.349..146G, 2006csxs.book..623T, 2013A&A...553A...7D}. 
\\
\\
The spectral and temporal characteristics of XRBs play a key role in their identification. LMXBs typically exhibit a softer X-ray spectrum with a photon index ($\Gamma$) in the range of 1 to 3 below 10 keV, while HMXBs are characterized by a harder X-ray spectrum, with $\Gamma$ values ranging from 1 to 2 above 10 keV \citep{2006csxs.book..623T, 2013A&A...553A...7D}. X-ray luminosities of XRBs generally fall within the range of L$_{X}$ =10$^{35}$$-$10$^{38}$ erg~s$^{-1}$; LMXBs in quiescence can exhibit luminosities below 10$^{34}$ erg~s$^{-1}$ \citep{2014ApJ...790..136V, 2018ApJ...853..150S}. X-ray color-color diagrams and X-ray luminosity functions (XLFs) serve as valuable tools for classifying XRBs as well. \citet{2003ApJ...595..719P} demonstrated that HMXBs and LMXBs exhibit distinct patterns in X-ray color-color diagrams. Numerous studies have noted that the XLFs of HMXBs and LMXBs follow distinct distributions, reflecting their association with different physical regions within a galaxy \citep{2003MNRAS.339..793G, 2004MNRAS.349..146G, 2012MNRAS.419.2095M, 2014MNRAS.437.1698M, 2019ApJS..243....3L, 2020ApJ...890..150C, 2021ApJ...912...31H, 2023MNRAS.519.4826A}.
\\
\\
Optical observations can play an important role in the classification of XRBs. Thanks to the {\it Hubble Space Telescope} ({\it HST}), detailed analyses of XRBs in nearby galaxies are made possible, particularly regarding their optical components. For instance, photometric observations are instrumental in constraining the mass of the companion star, which is a critical parameter to distinguish between HMXBs and LMXBs \citep{2021ApJ...912...31H}.
\\
\\
In recent years, multiwavelength machine learning techniques have become increasingly valuable for the classification of astrophysical sources in large X-ray catalogs \citep{2023ApJ...948...59C, 2023MNRAS.520.5065K, 2024ApJ...971..180Y}. By combining X-ray observations with photometric data from optical (e.g., {\it Gaia}), near-infrared (e.g., {\it 2MASS}), and mid-infrared (e.g., {\it WISE}) surveys, these methods enable the efficient identification of diverse source populations, including active galactic nuclei (AGN), XRBs, cataclysmic variables (CVs), young stellar objects (YSOs), and neutron stars (NSs). This multiwavelength approach has the capacity to enhance classification accuracy and also facilitates automated categorization of large samples, reducing the reliance on time-consuming spectroscopic follow-up. Tools such as MUWCLASS \citep{2022ApJ...941..104Y, 2016ApJ...821...54S, 2016ApJ...816...52H} exemplify this framework by providing a scalable and accessible classification of Chandra X-ray sources using photometric measurements, which makes them well suited for extragalactic X-ray population studies.\\
\\
In this study, we present the X-ray and optical properties of the XRBs in NGC 6946, a nearby spiral galaxy at a distance of 7.8 Mpc \citep{2018ApJ...860..117M}. X-ray properties such as X-ray spectra, light curves, color-color diagrams and XLFs of X-ray sources identified within the D$_{25}$ area in NGC 6946 were examined using the {\it Chandra} archival data spanning from 2001 September to 2017 June. In addition, we carried out a photometric analysis of the {\it HST}/ACS archival data of NGC 6946 to determine the optical counterparts of X-ray sources and to investigate properties of potential companion stars. While \cite{2008ApJS..177..465F} provided an initial inventory of X-ray sources in NGC 6946 based on five {\it Chandra} observations, their study lacked deep multiwavelength counterpart identification and spectral classification. In this work, we expand upon that foundational catalog by incorporating 16 years of {\it Chandra} data alongside {\it HST}/ACS optical imaging and multiwavelength machine learning classification, enabling a more detailed examination of the XRB population and its connection to the galaxy’s star formation history. X-ray and optical observations and analysis methods are given in section \ref{obs}. The results of X-ray and optical analysis are described in section \ref{result} and the summary in section \ref{sum}.

\section{Observations and Data Reduction}
\label{obs}

NGC 6946 was observed multiple times with {\it Chandra} over 16 years. Also, it has been observed with different {\it HST} cameras. We used the F435W, F606W and F814W data taken with the Advanced Camera of Surveys (ACS). The details of the observations used this study are given in Table \ref{obs-log}. 

\subsection{X-ray Observations}

Data reduction of the Chandra observations was performed using the {\sc ciao} \footnote{https://cxc.cfa.harvard.edu/ciao/} (Chandra Interactive Analysis of Observations) v4.12 software package together with the CALDB (Chandra Calibration Database) v4.9.3 calibration files. The X-ray sources are located on the ACIS-S3 (back-illuminated) chip in all Chandra observations. Reprocessed level-2 event files were generated using the {\sc chandra$\_$repro} script; subsequently the individual observations were reprojected onto a common astrometric reference frame and merged to produce a combined event file and exposure map covering the 0.3$-$8.0 keV energy band.\\
\\
Source detection was performed on the merged image using the {\sc wavdetect} task with wavelet scales of 2, 4, 8, and 16 pixels. The use of the merged dataset increases the sensitivity to faint sources by combining the effective exposure of all observations, while the broader range of wavelet scales improves the detection of both compact and moderately extended point-spread functions across the ACIS field of view. Source detection was restricted to the D$_{25}$ region of NGC 6946 (11$\farcm$2 $\times$ 10$\farcm$0; \cite{2002A&A...388....7W}) in order to minimize contamination from unrelated background sources.\\
\\
For each detected source, net counts were extracted from the individual observations in four energy bands: 0.3--1.0 keV (soft; S), 1--2 keV (medium; M), 2--8 keV (hard; H), and 0.3--8 keV (total; T). Only sources located within the D$_{25}$ region and having total net counts greater than 15 in the merged dataset were retained for subsequent spectral and population analysis. To ensure a robust source sample with reliable photometric measurements and hardness ratios, we adopted a more conservative selection criterion requiring total net counts greater than 15 in the merged dataset. This threshold minimizes the influence of statistical uncertainties associated with low-count sources while providing a reliable sample for the subsequent analysis. In addition, ten of these sources have previously been cataloged as SNRs \citep{2021ApJ...916...58K} and five sources overlap with foreground stars. We retained the remaining 98 sources as XRB candidates and investigated their classifications using the various analyses presented throughout this paper. The {\it Chandra} coordinates and properties of these sources are given in Table \ref{x-ray_par}. The energy spectra of these sources were obtained by extracting circular regions with a radius between 5$\arcsec$ $-$ 8$\arcsec$ around each source. Circular regions of the same size were selected from source-free areas located on the same chip as the sources to extract background counts. Source and background counts were obtained using the {\sc specextract} task in {\sc ciao} in the four energy ranges noted above. For the marginally detected sources, we extracted the upper limit in the observations using the {\sc srcflux} task at 90 \%  confidence level.
\\
\\
The energy spectra of the sources with counts $>$50 were obtained in the range of 0.3-8 keV by using the {\sc xspec} v.12.13.1 package (in {\sc heasoft} v6.32). These spectra were grouped with a minimum of 10 counts per bin using the {\sc grppha} task in the {\sc ftools} package \citep{1995ASPC...77..367B}. Power-law ({\scshape po}), a multi-color disk blackbody ({\scshape diskbb}) and {\scshape po + diskbb} models were used to fit the spectra. Absorption was included via two models: 'tbabs' and  Galactic absorption in which the value of N$_{H}$ was fixed at 2.18$\times$ 10$^{21}$ cm$^{-2}$ \citep{2016A&A...594A.116H} while the intrinsic value was left free. The best-fit model parameters of the XRBs which have $>$50 counts are given in Table \ref{best-fit}. The values in the table correspond to the model parameters applied to the C1 and C2 observations data, which have the longest exposure time. On the other hand, the unabsorbed flux values of all sources (see Table \ref{x-ray_par}) were derived from the {\sc webpimms} tool \footnote{https://cxc.harvard.edu/toolkit/pimms.jsp}. Using respective counts rates in the 0.3-8 keV, we determined flux values of each source in 0.3-8 keV assuming $\Gamma$=1.8 and Galactic column density N$_{H}$ = 2.18$\times$ 10$^{21}$ cm$^{-2}$.\\
\\
\citet{2008ApJS..177..465F} used five {\it Chandra} observations (C1$-$C5), totaling approximately 180 ks in exposure, to identify 90 X-ray sources in NGC 6946 with luminosities down to L$_{X}$ $\geq$ 10$^{36}$ erg~s$^{-1}$. Among these, five sources were classified as foreground stars, seven as supernova remnants (SNRs) based on optical and radio identifications, and four were associated with known supernovae. Their study focused primarily on the X-ray temporal properties of these sources, identifying 25 as long-term variable and 11 as transient. They constructed X-ray luminosity functions (XLFs) and found that a single power-law model best fit the data, with cumulative slopes of approximately 0.6$–$0.7. However, their analysis was limited to X-ray data alone, with no exploration of optical or infrared counterparts, and a shorter temporal baseline constrained their ability to detect long-term transients or variability trends.\\
\\
In contrast, our study extends this earlier work by incorporating eight Chandra observations spanning 16 years, with a combined exposure of approximately 250 ks, and by complementing the X-ray data with high-resolution {\it HST}/ACS optical imaging in the F435W, F606W, and F814W filters. This broader dataset enables a significantly improved census of the XRB population in NGC 6946. We adopt a consistent source detection threshold of $>$15 net counts and corresponding to a luminosity threshold of L$_X$~$\approx$~1~$\times$~10$^{37}$ erg~s$^{-1}$ (see Table \ref{x-ray_par} column (1) ) at the distance of NGC 6946.
\\ 
\\
For comparison with the (90-source) catalog of \citet{2008ApJS..177..465F}, we found that the 5 foreground stars and 8 SNRs in their list correspond to the foreground stars and SNRs identified in our classification. Therefore, these 13 sources were removed from further consideration, leaving a total 77 sources. In addition, after excluding the two sources located outside the D25 region, 75 sources from the Fridriksson et al. (2008) catalog remained for direct comparison with our final XRB candidate list. Of these 75 sources, 72 are present in our source list. The remaining three sources were not included in our final catalog because of our source-selection procedure. Two of these are very weak sources with count rates close to our adopted detection threshold and were therefore excluded from the final catalog. The third source is in a crowded region, very close to a known SNR. Therefore, it is difficult to reliably separate the X-ray emission from this source from that of the nearby SNR.\\
\\
Furthermore, our inclusion of astrometric corrections and multi-band photometry from {\it HST} enables the identification of optical counterparts and companion star mass estimates via CMD analysis, which were not addressed in the earlier study. We found that four sources classified as foreground stars in our catalog are also identified as foreground stars by \citet{2008ApJS..177..465F}. One additional source was classified as a foreground star in our catalog but is not listed as such by \citet{2008ApJS..177..465F}. Conversely, one source identified as a foreground star by \citet{2008ApJS..177..465F} was not classified as a foreground star in our catalog. Our foreground-star classification was based on positional matching between the astrometrically corrected X-ray source positions and their optical counterparts. Specifically, after applying the astrometric correction to the X-ray detections, we determined the corrected position of each X-ray source and examined the corresponding location in the optical images to identify possible foreground stars. For the source in question (i.e. source \#60 in Figure 4), the corrected X-ray position did not coincide with the optical source identified as a foreground star. Therefore, it was retained in our X-ray source sample. We also apply a supervised machine learning classification pipeline (MUWCLASS), which integrates optical, IR, and X-ray data to refine source types.

%we first excluded the 5 foreground stars identified in their catalog and 2 sources located outside the D$_{25}$ region considered in our analysis, leaving 83 sources. Of these 83 sources, 72 sources are present in our source list. Of the remaining 11 sources, 8 are present in our initial source list but were excluded from the final XRB sample because they are classified as SNRs. The remaining 3 sources were not retained in our final catalog because they are either marginal detections close to our adopted source-selection threshold or are located in a crowded region close to a known SNR. Overall, the differences between the two catalogs primarily reflect the different source-selection criteria and observational datasets used in the two studies.}

\subsection{Astrometry and {\it HST} Observations}

\begin{figure}
 \begin{center}
  \includegraphics[width=80mm]{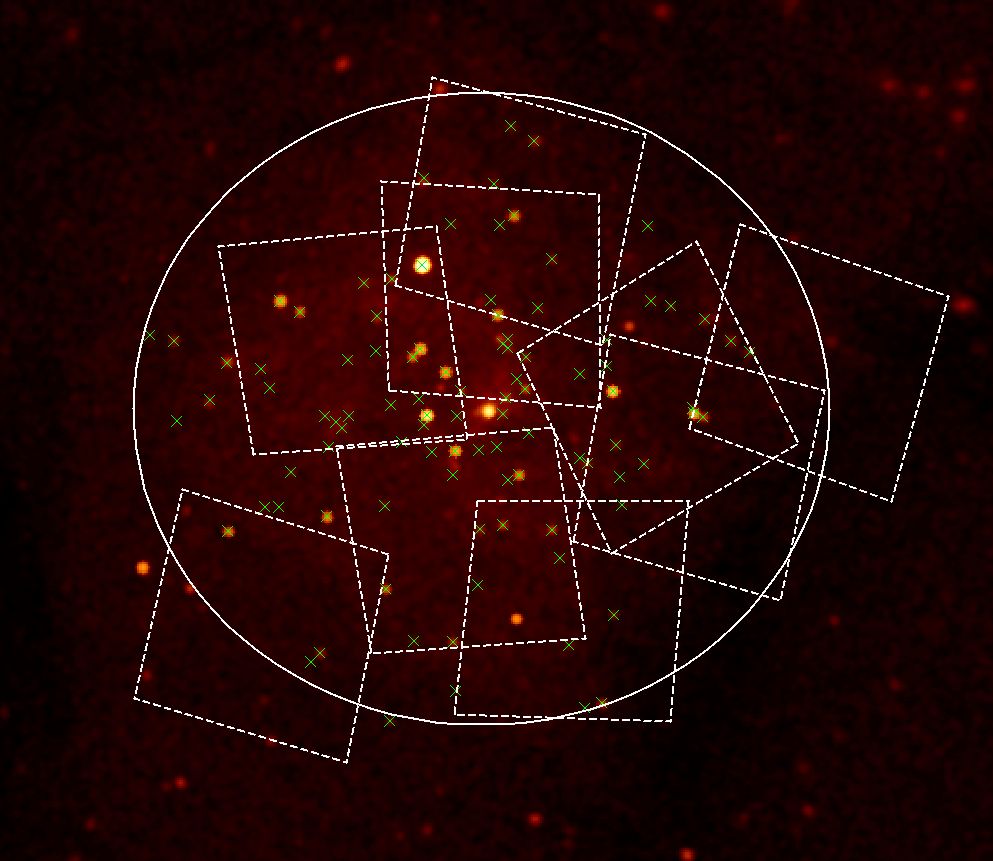}
 \end{center}
 \caption{Merged {\it Chandra} image of NGC 6946. {\it HST}/ACS observation fields used in this study are shown white-dashed squares and white ellipse represents D$_{25}$ region of NGC 6946. Detected XRB candidates are marked with green crosses. North is up.}
 \label{hst-ch}
\end{figure}

\noindent To study the optical properties of the optical counterpart candidates of the XRBs in NGC 6946, we used the {\it HST}/ACS observations covering 9 distinct fields. These fields are shown in Figure \ref{hst-ch}. The majority of XRB candidates (85 out of 98) fall into at least one of the {\it HST}/ACS fields shown in Figure \ref{hst-ch}. 
\\
\\
Relative astrometric correction is required between the {\it Chandra} and HST images to identify possible optical counterparts. We adopted the {\it Gaia} Early Data Release 3 (EDR3) catalog as the astrometric reference frame to compute the shift between the HST and {\it Chandra} images. First, we identified reference sources between the HST images and the {\it Gaia} catalog. A similar procedure was applied to identify reference sources between the {\it Chandra} images and the {\it Gaia} catalog. All X-ray sources identified in the {\it Chandra} observations and their counterparts in the {\it HST} F435W or F606W filter images were utilized. For each field, between four and six reference sources were identified. The correction resulted in an average shift in the source coordinates of 0$\farcs$1 in right ascension and 0$\farcs$16 in declination. After the coordinate correction, the uncertainty in the corrected position of each XRB candidate was estimated by combining the average astrometric shifts between  {\it HST}–{\it Gaia} and {\it Chandra}–{\it Gaia}, along with the X-ray positional uncertainty derived from the source counts. The resulting 1$\sigma$ error circles for each XRB candidate have radii ranging from 0$\farcs$27 to 0$\farcs$48. 
\\
\\
Depending on the source morphology—extended or point-like—we performed aperture or point spread function (PSF) photometry using the {\scshape DOLPHOT} v2.0 software. The data files (.flc and .drc) listed in Table~\ref{obs-log} were retrieved from the {\it HST} data archive\footnote{\url{https://mast.stsci.edu/search/ui/\#/hst}}. Standard data reduction procedures were applied to the .flc images following the steps outlined in the {\scshape DOLPHOT} manual \citep{2016ascl.soft08013D}. The tasks {\scshape acsmask} and {\scshape calcsky} were used to mask bad pixels in the .flc images and to generate sky images. We then ran {\scshape dolphot} to detect optical counterparts and perform photometry. The default {\scshape dolphot} zero points and aperture correction values were adopted.  
\\
\\
In many XRBs, the optical emission is often assumed to be dominated by the donor star, particularly during quiescence or the low-accretion phase. This assumption is based on the fact that the donor’s stellar photosphere typically outshines the accretion disk and other emission components in the optical wavelengths. Although the accretion disk can contribute significantly in the ultraviolet and X-ray bands, its optical contribution is usually less prominent when the mass transfer rate is low to moderate. As a result, the donor star’s spectral features and brightness can be reliably used to estimate its physical properties, such as mass and age, by comparing observed CMDs with theoretical isochrones. However, this assumption may not hold during high-accretion states or in systems hosting highly luminous accretion disks. In such cases, optical emission from the accretion disk and X-ray reprocessing can significantly enhance the optical brightness and shift the colors toward bluer values. This effect may lead to an overestimation of the donor mass inferred from CMDs and, consequently, introduce uncertainties in the classification of XRBs based on optical data alone. This effect is expected to be less significant for systems dominated by intrinsically luminous, massive donor stars. (see \cite{charles2006,casares2015,vanparadijs1995,hynes2005}). To constrain the masses of donor stars, CMDs of potential optical counterparts were constructed using the {\it HST} filters F435W, F606W, and F814W. Parsec isochrones\footnote{Padova and Trieste Stellar Evolution Code; http://stev.oapd.inaf.it/cgi-bin/cmd} given for the {\it HST}/ACS/WFC photometric systems with solar metallicity value were selected to create the CMDs. The distance modulus was calculated as 29.5 mag and the reddening was taken as E(B-V)=0.27 mag. \citep{2011ApJ...737..103S}.

\section{Results and Discussion}
\label{result}

\subsection{X-ray Variability}

\noindent The {\it Chandra} and {\it HST} observations used in this study, along with their observation IDs and filters, are listed in Table~\ref{obs-log}. The extracted unabsorbed fluxes, in the 0.3$-$8 keV energy band, are in the range 10$^{-16}-10^{-12}$ erg~cm$^{-2}$ s$^{-1}$. The corresponding X-ray luminosities fall in the range of 10$^{36}-10^{40}$ erg~s$^{-1}$ for an adopted distance of 7.8 Mpc for NGC 6946 (see Table \ref{x-ray_par}). Sources \#1, \#2, and \#5 (Table \ref{x-ray_par}) exhibit the highest luminosity in the sample and have previously been cataloged as ULXs \citep{2005ApJS..157...59L, 2019ApJ...881...38E}. Excluding these three ULXs, the highest L$_{X}$ among the remaining sources is a few $\times$ 10$^{38}$ erg~s$^{-1}$, consistent with the expected luminosity range of XRBs.
\\
\\
To examine the long-term X-ray variability of these XRB candidates, we used the variability parameter (S) and the maximum-to-minimum flux ratio ($F_{max}/F_{min}$). The variability parameter (S) is defined as S=($F_{max}-F_{min}$)/($\sigma_{max}^{2}+\sigma_{min}^{2})^{1/2}$ where $\sigma_{max}$ and $\sigma_{min}$ are the uncertainties in $F_{max}$ and $F_{max}$, respectively \citep{2019ApJ...879..112J}. Sources with $S > 3$ were classified as variable (v), while those with $F_{\mathrm{max}}/F_{\mathrm{min}} > 10$ were classified as transient (t) \citep{2017ApJ...836...50L, 2019ApJ...879..112J}. The calculated S values for the sources are given in Table \ref{x-ray_par} column (2).
\\
\\
For the 20 sources not detected in all observations, the $S$ parameter was calculated using their upper limit fluxes.
For 15 of these sources, the fluxes in the observations where they were detected are $\leq 10^{-16}$ erg~cm$^{-2}$~s$^{-1}$, and they show no evidence of variability. However, the fluxes of 8 out of the 20 sources vary by at least a factor of 10 across the observations in which they were detected, and these sources have already been classified as variable or transient. Among the sources, 40 exhibit long-term variability (S$>$3), 16 of which are transient sources (F$_{max}$/F$_{min}$$>$10). Including the three known ULXs (sources \#1, \#2 and \#5), a total of 12 sources show high variability (S$>$10). 

\subsection{X-ray Colors}

\begin{figure}
 \begin{center}
  \includegraphics[width=90mm]{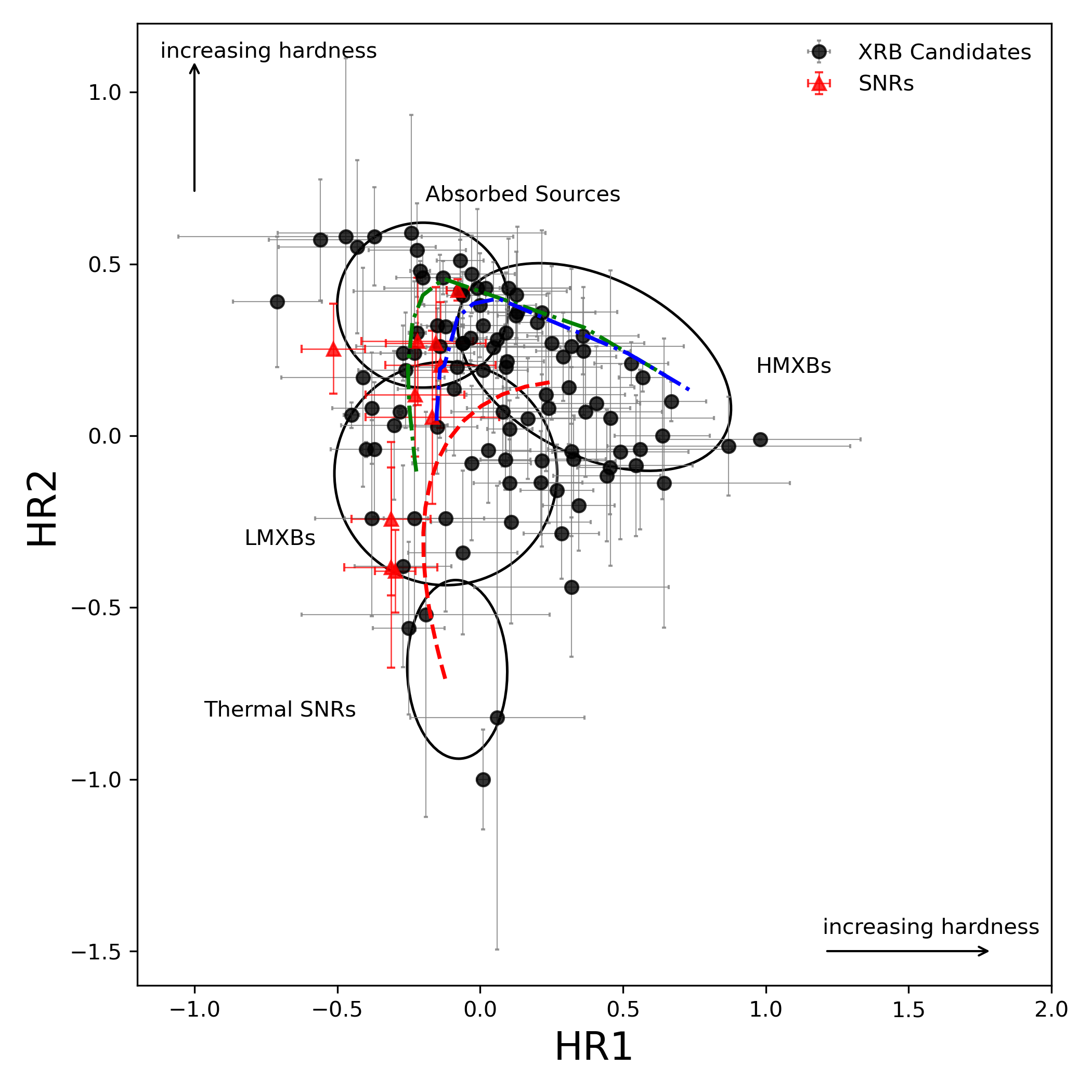}
 \end{center}
 \caption{The X-ray color-color diagram of all XRB candidates and known SNRs. Black circles represent all XRBs that are given Table \ref{x-ray_par} and red triangles are shown ten known SNRs. A simple power-law spectral model with increasing $\Gamma$ from 0.3$-$3.5 is shown with red dashed line. A single power-law model with photon index value of 1.5 and 2.0 and added increasing N$_{H}$ are shown with blue and green dashed-dot lines, respectively. Regions (indicated as solid black circles and ellipses) encompassing LMXBs, HMXBs and SNRs etc. are taken from \cite{2003ApJ...595..719P}. Note that there is considerable overlap in the data points making the CCD an unambiguous classification tool in this case.}
 \label{c-c}
\end{figure}

\noindent Spectral fitting was not possible for observations with insufficient photon statistics caused by short exposures and/or faint sources. In such cases, we constructed a color–color diagram (CCD) following the procedure described by \cite{2003ApJ...595..719P}. CCDs are particularly effective in distinguishing between soft thermal emission, often originating from the accretion disk, and harder, non-thermal emission associated with the corona or relativistic jets. 
\\
\\
We determined the CCD based on hardness ratios calculated for each observation as HR$_{1}$=(H-M)/T and HR$_{2}$=(M-S)/T, respectively, where S is counts in 0.3$-$1 keV, M is counts in 1$-$2 keV, H is counts in 2$-$8 keV and T is the total counts in 0.3$-$8 keV energy bands. A plot of HR$_{1}$ vs HR$_{2}$, for all XRB candidates, is shown in Figure \ref{c-c}. To examine the behavior of sources with different power-law spectra, we generated HR tracks by varying the powerlaw index in the range 0.3$-$3.5. The resulting track is overlaid on the CCD (as a red dashed line). In addition, the power-law model tracks with increasing hydrogen column density, assuming a constant photon index value of 1.5 and 2.0, respectively are also shown on the CCD (as blue and green dashed-dot lines respectively). Particular regions (indicated as solid black circles and ellipses) encompassing LMXBs, HMXBs and SNRs etc are taken from \cite{2003ApJ...595..719P}. As seen in Figure \ref{c-c}, most of the sources (58$\%$) are consistent with HMXBs and approximately one-third of the sources are in agreement with LMXBs. When the count-rate uncertainties are taken into account, the source distribution in the CCD becomes broader, particularly for faint sources, increasing the overlap between the regions associated with different source classes. Consequently, classifications based solely on X-ray colors should be interpreted with caution. In this work, the CCD is therefore used as a diagnostic rather than a definitive classification tool. The final classifications are established by combining the X-ray color information with multiwavelength counterpart identifications and the MUWCLASS machine-learning classifications, providing a more robust characterization of the X-ray source population.

\subsection{X-ray Luminosity Function}

\begin{figure*}
 \begin{center}
  \includegraphics[width=180mm]{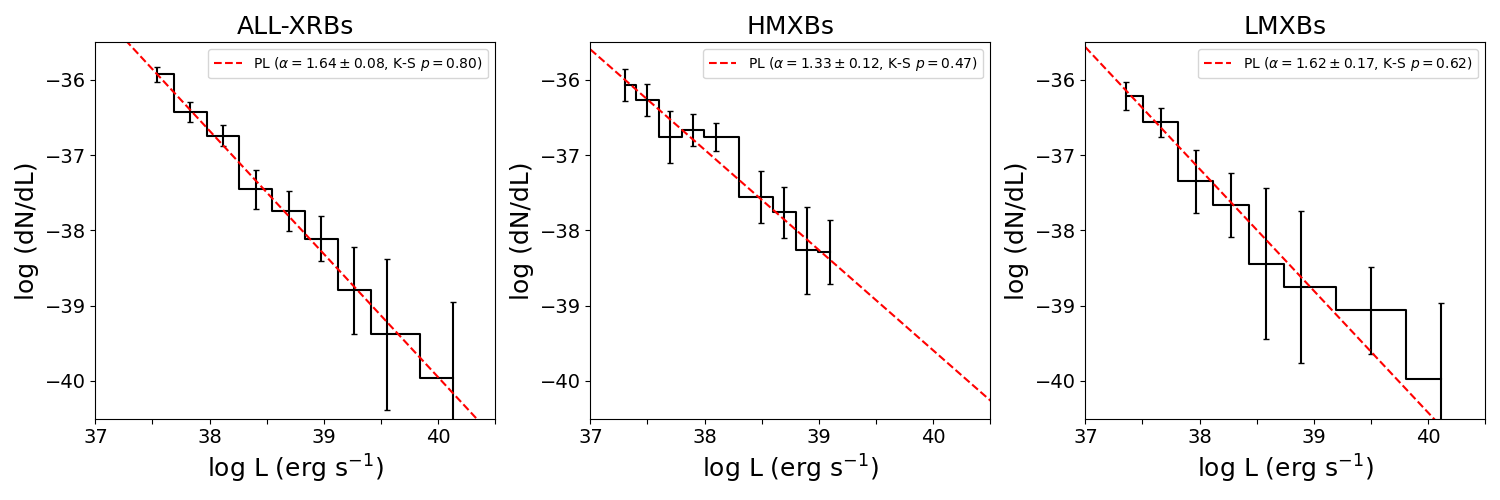}
 \end{center}
 \caption{The differantial XLFs for all XRB candidates (98 X-ray sources in our list, including ULXs), HMXBs ( 57 sources classified as HMXBs in Table \ref{x-ray_par}) and LMXBs ( 28 sources classified as LMXBs in Table \ref{x-ray_par}) which are classified based on their X-ray data. The red dashed lines indicate the best-fit single power-law models obtained from an unbinned maximum likelihood estimation (MLE). The goodness of fit was evaluated using a Monte Carlo Kolmogorov–Smirnov (K–S) test.}
 \label{xlf}
\end{figure*}

\noindent X-ray luminosity functions (XLFs) provide valuable information on the population statistics of XRBs by describing how the number of sources varies with luminosity. They also help distinguish populations in different environments, such as the galactic field versus globular clusters, and enable estimates of the total X-ray luminosity contributed by a population. Indeed, many studies showed that the XLFs of HMXBs scale with the star formation rate (SFR) of the host galaxy and the XLFs of LMXBs scale with stellar masses \citep{2003MNRAS.339..793G, 2004MNRAS.347L..57G, 2004MNRAS.349..146G, 2012MNRAS.419.2095M, 2014MNRAS.437.1698M, 2019ApJS..243....3L}. 
\\
\\
The XLF is described by a power-law, which reflects the underlying physical processes governing the population. The power-law is typically expressed as:\\
\\
N($L_{X}$) $\propto$ $L^{-\alpha}_{X}$\\
\\
\noindent where N($L_{X}$) is the number of sources with luminosity L$_{X}$, and the index $\alpha$ determines the slope of the function. \cite{2020ApJ...890..150C} examined the XLFs of XRBs in the M101 galaxy and found that XLFs for HMXBs and LMXBs exhibit different power-law slopes. The XLF of HMXBs extends to higher luminosities and is flatter than the XLF of LMXBs \citep{2003ChJAS...3..257G, 2012MNRAS.419.2095M, 2020ApJ...890..150C}. In some cases, the XLF of XRBs cannot be adequately described by a single power-law and a broken power-law is used instead with the introduction of an additional index representing a change in the slope of the function at some threshold break luminosity L$_{break}$. This luminosity often indicates a transition between different types of sources and/or changes in accretion modes.
\\ 
\\
We constructed the XLFs of sources classified as LMXBs and HMXBs based on their X-ray color–color values. We fitted the unbinned X-ray luminosity functions using single and broken power-law models. The model parameters were estimated using an unbinned maximum likelihood estimation (MLE) approach. The likelihood optimization was performed in Python using the {\sc scipy.optimize} package \citep{2020NatMe..17..261V}. The goodness of fit was assessed using a Monte Carlo Kolmogorov–Smirnov (K–S) test implemented with the {\sc scipy.stats} package. Uncertainties on the best-fit parameters were estimated from 5000 bootstrap resamplings, and the corresponding 68\% and 95\% confidence intervals were derived from the bootstrap distributions. We found that a single power-law provides the best fit to the data, while a broken power-law model yielded a poor fit. 
\\
\\
Figure \ref{xlf} presents the differential XLFs of all XRB candidates, HMXBs, and LMXBs. The XLFs were constructed using the average X-ray luminosities of the sources. We fitted the distributions with both single and broken power-law models to test for the presence of a statistically significant break. The break luminosity and the corresponding slopes could not be robustly constrained, and the fit did not provide a statistically significant improvement over the single power-law model. Therefore, the XLFs are best described by a single power-law model over the sampled luminosity range. The similarity between the XLFs of the total XRB sample and the HMXB population suggests that HMXBs make a substantial contribution to the overall X-ray source population in this galaxy.
\\
\\
Subsequently, we constructed the XLFs for LMXBs, IMXBs, and HMXBs classified via their optical counterparts, as described in Section~\ref{opt_count}. We fitted the distributions with both single and broken power-law models. For all three subclasses, the single power-law model provides a more statistically stable and reliable description of the data. In contrast, the broken power-law model shows substantial uncertainties, especially in the high-luminosity (bright-end) slope. The results suggest that the X-ray-based classification is statistically more reliable for HMXBs, whereas the optical-based classification appears to better constrain the LMXB population.

\subsection{Optical Counterparts}
\label{opt_count}

\noindent Optical observations provide key insights into the properties of the companion star, including its type, mass, and orbital parameters. In HMXBs, the dominant source of optical emission is the massive, hot companion star, which emits strongly in the optical and ultraviolet bands. In LMXBs, however, the optical emission primarily comes from the accretion disk surrounding the compact object, where material from the companion star is heated as it spirals inward. Additionally, X-rays produced near the compact object can be reprocessed by the outer parts of the accretion disk or the surface of the companion, contributing to the optical light \citep{1995CAS....26.....L, 2002apa..book.....F, 2006csxs.book..215C, 2006MNRAS.371.1334R}. 
\\
\\
After the astrometric correction between {\it Chandra} and {\it HST} images, the possible optical counterparts of 57 out of 85 XRB candidates were identified. Of these sources, 35 have a single optical counterpart, while 22 sources have multiple optical candidates within the error region (see Figure \ref{opt_counterparts}). The de-reddened magnitudes and color values of possible optical candidates are given in Table \ref{opt_par}. CMDs enable us to understand the physical properties and evolutionary stages of donor stars in XRBs. They are particularly valuable in XRB research because they help distinguish donor star types and reveal the effects of accretion processes on the optical emission. Assuming that the optical emission is dominated by the donor star, the CMD for the identified optical candidates of XRBs in NGC 6946 is shown in Figure \ref{cmd}. Overlaid on the figure are PADOVA mass isochrones corresponding to $5M_{\odot}$, $8M_{\odot}$, $20M_{\odot}$ and $30M_{\odot}$. Some optical counterparts lie between $5,M_{\odot}$ and $8,M_{\odot}$, while others exceed $20,M_{\odot}$. In addition, 40 sources have been classified as HMXBs and 17 as IMXBs. The remaining sources are classified as LMXBs, assuming that their donor stars were not detected in the {\it HST} observations because of their low masses ($< 3,M_{\odot}$).

\subsection{Infrared Counterparts}

\noindent The primary sources of infrared emission from XRBs are the donor star, the accretion disk and, in some cases, relativistic jets. Infrared observations provide information on mass transfer, disk–jet coupling, and the physical properties of donor stars in XRBs. Furthermore, the reduced absorption of infrared light by dust and gas allows the study of XRBs located in dense, obscured regions. 
\\
\\
The near-infrared counterparts of 16 sources in NGC 6946 were found from the 2MASS All-Sky Catalog of Point Sources \citep{2003yCat.2246....0C}. The color–magnitude diagram constructed using the J and K band magnitudes is shown in Figure \ref{j-k}. \citet{2016A&A...586A..64S} demonstrated that foreground stars, AGNs, XRBs occupy distinct regions in the near-infrared color–magnitude diagram they constructed for X-ray sources in the Draco dwarf spheroidal galaxy. Sources with K = 15.5 (and brighter) and J-K $\geq$ 1 lie in the AGN region (shown as dashed black lines in Figure \ref{j-k}). 
When considering the associated errors, four sources (\#3, \#20, \#21, and \#50) lie in or near the AGN region. Sources \#3 and \#21 are not included in our list as they were previously classified as SNRs \citep{2021ApJ...916...58K}. 
Source \#20, which appears in the absorption source region in the X-ray color-color diagram, appears to be located in a very crowded field based on {\it HST} images. \#50 is located in the LMXB region based on X-ray CCD classification and no optical counterpart detected in the error circle on {\it HST} image. Given the average positional uncertainty of 2MASS (~3$\arcsec$; \cite{2016A&A...586A..64S}), the brightness of this source may have been affected by contributions from its surrounding environment. Figure \ref{j-k} illustrates that sources \#10, \#16, \#70, \#93 and \#101 are classified as foreground stars. A detailed examination of the {\it HST} images reveals that these sources spatially coincide with stellar objects. Although source \#32 is classified as an LMXB in the X-ray color-color diagram, it does not fall into the appropriate region in Figure \ref{j-k}. Additionally, the absence of {\it HST} observations for this source limits the available information about it. The remaining six sources (\#11, \#39, \#47, \#51, \#60, \#71, ) are classified as XRBs, consistent with previous classifications.

\subsection{Classification based on Machine Learning}
\label{muwclass}

As an additional approach, we employed supervised machine learning to classify the X-ray sources detected in NGC 6946.  
The goal was to identify the potential nature of these sources by utilizing their multiwavelength photometric properties across the optical, infrared, and X-ray bands. For this purpose, we employed the publicly available MUWCLASS pipeline (MultiWavelength CLASSification). It is specifically designed to classify sources from the {\it Chandra} Source Catalog (CSC 2.0) using optical and infrared data \citep{2022ApJ...941..104Y}. While MUWCLASS was trained on the CSC 2.0 dataset, which includes a variety of source types, we acknowledge that the model’s domain may not exactly match that of an extragalactic, star-forming system like NGC 6946. As such, we treat the ML classification as supplementary and not necessarily definitive.\\
\\
We initiated the analysis by performing a cone search query centered at the galaxy's coordinates with a radius of 10$\arcmin$ using the CSC 2.0 service. A total of 148 sources with signal-to-noise ratios (S/N) $\geq$ 3 were selected for further analysis. Each source was then cross-matched with external multiwavelength catalogs: Gaia EDR3 (G, BP, RP magnitudes), 2MASS (J, H, K), WISE, AllWISE (W1, W2, W3, W4). Cross-matching was performed using the {\sc astroquery} module with a matching radius of up to 10$\arcsec$. \\
\\
For training, we utilized the \texttt{CSC$\_$TD$\_$MV$\_$remove.csv} file provided with the MUWCLASS code. This dataset comprises labeled X-ray sources from the {\it Chandra} Source Catalog with confirmed classifications, including LMXBs, HMXBs, cataclysmic variables (CVs), young stellar objects (YSOs), neutron stars (NSs), AGNs, low-mass stars (LM-Stars), and high-mass stars (HM-Stars). The classification was performed using a Random Forest classifier, trained with scikit-learn's implementation (RandomForestClassifier) and standardized via StandardScaler. Despite the limited number of features, the model achieved reasonable discrimination across source types based on mid-infrared characteristics. The final MUWCLASS classification of the sources in our sample, along with their probabilities, is presented in Table \ref{x-ray_par} column (6). We only list sources with confidence level $>$70\%, resulting in 23 AGNs, 3 NSs, 4 LMXBs, and \textbf{one} low-mass star. Sources with a lower threshold are indicated with a question mark. \\
\\
It is generally recognized that ML classification requires a large and balanced training database and the precision of the known parameters on which classifications is based should be very good otherwise the uncertainties in the ML classification can be significant and produce biased results. This can lead to accuracies of the classification as low as 70-80\%. We note that the training dataset of \cite{2024ApJ...971..180Y}, contains $\sim$1500 AGNs, $\sim$30 HMXBs, and $\sim$50 LMXBs, i.e., the database is biased towards AGN, which is likely to produce unbalanced results. However, as noted elsewhere the expected contribution from background AGN is at the $\sim$15-22\% level: the ML misclassification is of the same order.

\section{Summary and Conclusion}
\label{sum}
In this study, we performed a comprehensive classification of X-ray sources based on their X-ray and optical properties in the star-forming galaxy NGC 6946. Utilizing eight archival {\it Chandra} observations, we analyzed the temporal and spectral characteristics of \textbf{98} X-ray sources detected in the 0.3–8 keV energy range. Optical counterparts were identified and examined using archival {\it HST}/ACS imaging. In addition to traditional classification methods, we employed a multiwavelength machine learning approach using the MUWCLASS tool, which integrates photometric data from X-ray, optical ({\it Gaia}), near-infrared ({\it 2MASS}), and mid-infrared ({\it WISE}) surveys. 
\\
\\
\noindent We summarize our findings as follows:
\begin{itemize}

  \item We identified 113 point-like X-ray sources with net counts greater than 15 in the Chandra observations of NGC 6946. Ten of these sources have previously been cataloged as SNRs, while an additional five were identified as foreground stars and excluded from further analysis. The remaining 98 sources were therefore considered XRB candidates, with luminosities in the range of L$_{\rm X} \sim 10^{36}$–$10^{40}$ erg~s$^{-1}$. Among these, three sources have been identified as ULXs in previous studies.

  \item We estimated the contribution of background AGN using CXB (Cosmic X-ray Background) logN–logS relations at our flux limit (S$_{lim} \sim 10^{-15}$ erg~ cm$^{-2}$ s$^{-1}$) in the 0.3–8 keV band. Using published CXB source counts this flux limit implies an AGN surface density of approximately 400$-$600  deg$^{-2}$ \citep{2003ApJ...588..696M, 2008MNRAS.388.1205G}. Given the D$_{25}$ area of the galaxy ($\sim$0.024 deg$^{2}$) we therefore expect $\sim$12$-$15 background AGN within the D$_{25}$ region. 
  %Within the D$_{25}$ region, we detect \textbf{98 X-ray sources after excluding previously cataloged supernova remnants and foreground stars.} 
 
  \item To examine the variability of the sources, we calculated the variability significance parameters ($S$) and flux ratios ($F_{\rm max}/F_{\rm min}$). A total of 40 sources were classified as variable ($S > 3$), and among these, 16 were identified as transient sources based on their high variability amplitudes ($F_{\rm max}/F_{\rm min} > 10$).

  \item An X-ray CCD was constructed to distinguish between HMXBs and LMXBs. Based on their spectral characteristics, out of the 98 sources, 57 were classified as HMXBs and 28 as LMXBs. Of the remaining 13 sources six appear to be heavily absorbed and the nature of the remaining seven is unclear.

  \item Using the stellar mass and star formation rate of NGC 6946, we estimated the expected numbers of high-mass and low-mass X-ray binaries based on well-established scaling relations \citep{2004MNRAS.349..146G}. The expected numbers of LMXBs (8–12) and HMXBs (30–44) inferred from these scaling relations indicate a substantially larger HMXB population than LMXB population, consistent with the trend seen in our color-color diagram classification, where we identify 28 LMXBs and 57 HMXBs.

  \item After applying astrometric corrections, 35 X-ray sources were found to have a single optical counterpart, while 22 sources were associated with multiple potential counterparts. CMDs were constructed for these sources to estimate the mass of the donor stars. Based on their position in the CMD, 40 sources with donor masses greater than 8~M${\odot}$ were classified as HMXBs, and 17 sources with donor masses in the range of 5–8~M${\odot}$ were identified as IMXBs. No sources with donor masses below 3~M$_{\odot}$ were detected. However, assuming that the donor stars of the 28 sources lacking optical counterparts are too faint (i.e., low mass) to be detected with {\it HST}, these were classified as LMXBs. The high fraction of HMXBs (both X-ray and optically classified) is consistent with the star formation rate ($\sim$5–10 M$_\odot$ yr$^{-1}$) of NGC 6946.

  \item The differential XLFs were constructed for all XRBs, HMXBs, and LMXBs based on their X-ray properties, and the single power-law model provided the best fit for all three XLFs. The best-fit slopes are  $\alpha$=1.64±0.08 for all XRBs,  $\alpha$=1.33±0.12 for HMXBs, and $\alpha$=1.62±0.17 for LMXBs. The HMXB luminosity function appears somewhat shallower than that of the LMXBs, suggesting a relatively larger fraction of luminous HMXBs. However, the slope values remain consistent within their statistical uncertainties. The slope obtained for the HMXB population is broadly consistent with previous studies of star-forming galaxies, supporting the view that the X-ray source population of NGC 6946 is dominated by high-mass X-ray binaries.

  \item Although the differential XLFs constructed from the optically classified HMXB, IMXB, and LMXB populations are based on smaller source samples and therefore have lower statistical significance, all three luminosity functions are adequately described by a single power-law model. The best-fit slopes are $\alpha$=1.53±0.14 for HMXBs, $\alpha$=1.57±0.37 for IMXBs, and $\alpha$=1.52±0.09 for LMXBs. Within the statistical uncertainties, these slopes are mutually consistent and are also broadly consistent with those derived from the X-ray-based classification.

  \item We applied a multiwavelength machine learning approach to classify X-ray sources in NGC 6946, identifying a dominant population of AGNs and XRBs.  Based on confidence level $>$70\% the classification yielded four LMXBs, three neutron stars, 23 AGNs, and one low-mass stars. The AGN number is inconsistent with the estimated value (noted above), potentially indicating the challenge of combining multiwavelength data from instruments with limited spatial and spectral resolution. 

  \item To evaluate the effect of background AGN contamination, we recomputed the XLF after excluding the sources classified as AGN by the machine-learning algorithm. The resulting XLF is well described by a power law with a slope of $\alpha$=1.58±0.09, compared to $\alpha$=1.64±0.08 obtained using the complete source sample. The two slopes are consistent within their statistical uncertainties, indicating that the exclusion of AGN candidates does not significantly affect the derived XLF. This suggests that the luminosity function is dominated by the intrinsic X-ray binary population.
  
  \item The ratio of HMXBs to LMXBs in NGC 6946 provides insight into the galaxy’s recent star formation activity. Based on our classification methods, we find that approximately 57 sources are consistent with HMXBs and 28 with LMXBs, yielding an HMXB-to-LMXB ratio of $\sim$2.0, depending on the classification method. This ratio is comparable to those observed in other actively star-forming galaxies. For instance, in M101, a late-type spiral with a moderate SFR ($\sim$2–3 M$_\odot~yr^{-1}$), \citet{chandar2020} reported an HMXB-dominated population with a similar HMXB-to-LMXB ratio of $\sim$2. In the Antennae galaxies — a prototypical starburst merger system with an SFR $>$ 10 M$_\odot yr^{-1}$ — \citet{zezas2002} and \citet{mineo2012} found an even higher dominance of HMXBs, with the ratio exceeding 3.
  
\end{itemize}

  The HMXB prevalence in NGC 6946 is consistent with its elevated star formation rate ($\sim$5–10 M$_{\odot}$ yr$^{-1}$), inferred from H$\alpha$ and far-infrared measurements (e.g., \citet{murphy2011}), and reinforces the empirical correlation between star formation activity and the production of short-lived, massive XRBs \citep{grimm2003,mineo2012}. The presence of numerous massive donor candidates in the CMD, particularly in the panel constructed using the F435W filter, is therefore expected for a galaxy with ongoing star formation. The blue-sensitive filter preferentially traces young, massive stellar populations, naturally leading to an enhanced number of high-mass systems in the optical sample. These findings support the interpretation that NGC 6946 is currently in an XRB population phase dominated by young, massive binary systems.

The classification of X-ray sources in NGC 6946 benefits from a multi-criteria approach that combines X-ray properties with multiwavelength counterpart identifications (including HST and 2MASS data) and the independent MUWCLASS machine-learning classifications. Although the X-ray CCD provides a useful diagnostic for investigating the nature of X-ray sources, the relatively large color uncertainties for many detected sources prevent an unambiguous classification based solely on X-ray colors. By combining these complementary diagnostics, this work provides a more reliable and robust characterization of the X-ray source population in NGC 6946 than could be achieved using any individual method alone.

%% Please use the acknowledgment and contribution environments. This will 
%% be anonomyized when the "anonymous" style option is used. 
\begin{acknowledgments}
The authors thank the anonymous referee for the constructive comments and suggestions that improved this manuscript. This work was supported by the project numbered 122C042 under the 2218-National Postdoctoral Research Fellowship Program, conducted by the Scientist Support Programs Presidency (BİDEB) of the Scientific and Technological Research Council of Türkiye (TÜBİTAK). This paper employs a list of Chandra datasets, obtained by the Chandra X-ray Observatory, contained in the Chandra Data Collection ~\dataset[DOI:10.25574/cdc.551]{https://doi.org/10.25574/cdc.551} . 
\end{acknowledgments}

\begin{table}
	\centering
 	\caption{The list of archival observations used in this work. C and H labels represent {\it Chandra} and {\it HST} observations, respectively.}
	\label{obs-log}
	\begin{tabular}{ccccccr} % four columns, alignment for each
		\hline  \hline       
 Label & ObsID (Filter) & Date & Exp.  \\
  & & & ks \\ 
  \hline
  C1 & 1043 & 2001-09-07 & 60 \\
  C2 & 4404 & 2002-11-25 & 30 \\
  C3 & 4631 & 2004-10-22 & 30 \\
  C4 & 4632 & 2004-11-06 & 30 \\
  C5 & 4633 & 2004-12-03 & 30 \\
  C6 & 13435 & 2012-05-21 & 20 \\
  C7 & 17878 & 2016-09-28 & 40 \\
  C8 & 19040 & 2017-06-11 & 10 \\

\hline
 H1 & JD9G01010 (F435W) & 2016-10-28 & 5.61 \\
    & JD9G01020 (F606W) & 2016-10-28 & 4.47 \\
 H2 & JD9G02010 (F435W) & 2016-11-02 & 5.61 \\
    & JD9G02020 (F606W) & 2016-11-02 & 4.48 \\
 H3 & JD9G03010 (F606W) & 2016-10-26 & 2.43 \\
    & JD9G03020 (F814W) & 2016-10-26 & 2.57 \\
 H4 & JDK409020 (F435W) & 2018-12-10 & 2.70 \\
    & JDK409010 (F606W) & 2018-12-10 & 2.57 \\
 H5 & JDK406010 (F606W) & 2019-01-26 & 2.53 \\
    & JDK406020 (F814W) & 2019-01-26 & 3.71 \\
 H6 & JE3Y06020 (F435W) & 2020-11-11 & 6.41\\
    & JE3Y06010 (F606W) & 2020-11-11 & 1.66 \\
 H7 & JE3Y11020 (F435W) & 2020-11-13 & 6.41\\
     & JE3Y11010 (F606W) & 2020-11-13 & 1.66 \\
 H8 & JE3Y13020 (F435W) & 2020-11-07 & 6.41 \\
    & JE3Y13010 (F606W) & 2020-11-06 & 1.66 \\
 H9 & JE3Y58020 (F435W) & 2021-05-04 & 6.19 \\
    & JE3Y58010 (F606W) & 2021-05-04 & 1.66 \\
    \hline
	\end{tabular}
\end{table}

\begin{longrotatetable} 
\begin{deluxetable*}{cccccccccccccccccc}
\tabletypesize{\scriptsize} 
\tablewidth{\textwidth}
\tablecaption{X-ray luminosity and properties of XRB candidates identified in NGC 6946: (-) The source is located outside the image field. (1) X-ray luminosity obtained by combining eight {\it Chandra} observations; values are in units of 10$^{38}$ erg~s$^{-1}$, (2) S: variability parameters \protect\citep{2019ApJ...879..112J}, (3) the ratio of the maximum and minimum flux in 0.3-8 keV, (4) v: variable sources, t: transient sources, n: none-variable sources, (5) Source classification determined from X-ray color-color values; H: High-mass XRBs, L: Low-mass XRBs, A: absorbed sources and uc: unclear, (6) Source classification based on the MUWCLASS pipeline. Abbreviations used here are defined in Section \ref{muwclass} and the class probabilities are given in parentheses. "?" denotes cases with less confident ($<$70\% confidence) classifications, (7) 'y' : optical counterpart(s) observed, 'n' : optical counterpart(s) not observed. \label{x-ray_par}}
\tablehead{
\colhead{srcID} & \colhead{R.A.} & \colhead{Decl.} & \multicolumn{8}{c}{L$_{X}$ ($\times$10$^{38}$ erg~s$^{-1}$)} & \colhead{L$_{X(combined)}$} & \colhead{S} & \colhead{ratio} & \colhead{var.}  & \multicolumn{2}{l}{Class} & \colhead{optical det.} \\ & \colhead{deg}& \colhead{deg}& \colhead{1043} & \colhead{4404} & \colhead{4631} & \colhead{4632} & \colhead{4633}& \colhead{13435} & \colhead{17878} &\colhead{19040}	&\colhead{(1) } &\colhead{(2)} &\colhead{(3)}& \colhead{(4)} &\colhead{(5)} &\colhead{(6)} & \colhead{(7)}
}
\startdata
1	&	308.75308 & 60.19187 & 93.37 &	82.06	&	85.97	&	106.27	&	103.01	&	71.19	&	73.14	&	84.31	& 187.39    &	14.67	&	1.49	&	v	&	L	&  ?             & n  \\
2	&	308.75055 & 60.15225 & 21.27 &	20.26	&	14.03	&	14.11	&	13.56	&	7.80	&	9.01	&	10.29	& 33.18     &	18.58	&	2.73	&	v	&	L	&  L (0.83)      & y  \\
4	&	308.60886 & 60.15263 & 9.72	 &	4.30	&	-       &	-	    &	2.83	&	7.79	&	-	    &	3.88	& 15.75     &	15.63	&	3.44	&	v	&	H	&  ?             & y  \\
5	&	308.65216 & 60.15848 & 9.84	 &	14.90	&	-	    &	10.79	&	11.64	&	6.93	&	-	    &	10.92	& 24.45     &	10.46	&	2.15	&	v	&	L	&  L (0.79)      & y  \\
6	&   308.73541 & 60.14284 & 4.52	 &	3.31	&	3.94	&	4.39	&	6.09	&	4.73	&	6.18	&	4.79	& 9.83      &	6.03	&	1.86	&	v	&	H	&  ?             & n  \\
7	&   308.75401 & 60.16975 & 4.47	 &	4.42	&	6.92	&	5.37	&	5.38	&	3.48	&	6.00	&	5.23	& 10.70     &	6.04	&	1.99	&	v	&	H	&  ?             & y  \\
8	&	308.74056 & 60.16360 & 4.17	 &	2.63	&	2.96	&	2.93	&	4.51	&	5.32	&	6.12	&	6.00	& 8.65      &	7.61	&	2.33	&	v	&	H	&  ?             & n  \\
9	&	308.82822 & 60.18238 & 5.14	 &	2.81	&	3.90	&	5.14	&	4.90	&	-	    &	10.30	&	4.64	& 11.15     &	13.29	&	3.66	&	v	&	H	&  ?             & y  \\
%10	&	308.70337 & 60.09859 & 3.75	 &	2.68	&	2.69	&	3.24	&	4.14	&	3.91	&	1.12	&	0.39	& 6.39      &	7.53	&	3.70	&	v	&	L	&  LM-Star (0.94)& n  \\
11	&	308.70439 & 60.20488 & 2.76	 &	2.92	&	2.92	&	2.30	&	3.30	&	1.50	&	-	    &	2.51	& 5.92      &	4.35	&	2.20	&	v	&	H	&  ?             & y  \\
12	&   308.70190 & 60.13655 & 2.38	 &	1.95	&	3.50	&	1.29	&	3.44	&	0.48	&	3.73	&	1.22	&  5.10     &	9.54	&	7.77	&	v	&	H	&  L (0.77)      & y  \\
13	&	308.60413 & 60.15170 & 2.30	 &	0.05	&	-	    &	-	    &	0.19	&	0.02	&	-	    &	2.40	& 2.68      &	4.55	&	142.64	&	t	&	H	&  ?             & y  \\
14	&	308.80330 & 60.12553 & 2.15	 &	1.27	&	2.57	&	3.16	&	3.87	&	2.52	&	1.92	&	1.76	& 4.95      &	6.56	&	3.05	&	v	&	H	&  L (0.72)      & n  \\
15	&	308.75809 & 60.16768 & 1.98	 &	1.62	&	2.25	&	1.69	&	2.45	&	1.27	&	1.05	&	1.65	& 3.78      &	4.26	&	2.33	&	v	&	H	&  ?             & y  \\
%16	&	308.64357 & 60.17582 & 1.20	 &	1.23	&	-	    &	1.05	&	2.51	&	1.19	&	-	    &	2.87	& 3.21      &	3.03	&	2.74	&	v	&	L	&  ?             & n  \\
17	&	308.77223 & 60.10644 & 1.21	 &	0.94	&	1.54	&	1.49	&	1.51	&	1.66	&	1.04	&	1.13	& 2.70      &	2.45	&	1.76	&	n	&	H	&  ?             & y  \\
18	&	308.60314 & 60.17740 & 0.75	 &	0.47	&	-	    &	-	    &	0.09	&	0.22	&	-	    &	0.46	& 1.15      &	5.56	&	8.40	&	v	&	H	&  NS (0.69)     & y  \\
19	&	308.76926 & 60.18818 & 0.78	 &	0.87	&	0.49	&	0.77	&	0.93	&	0.09	&	0.75	&	0.44	& 1.48      &	4.20	&	10.87	&	t	&	L	&  ?             & y  \\
20	&	308.57946 & 60.16886 & 0.74	 &	0.16	&	-	    &	- 	    &	0.03	&	0.40	&	-	    &	1.37	& 1.11      &	3.28	&	50.67	&	t	&	A	&  ?             & y  \\
22	&	308.63543 & 60.13940 & 0.62	 &	0.01	&	-	    &	-	    &	0.00	&	0.48	&	-	    &	0.00	&  0.96     &	7.15	&	45.06	&	t	&	H	&  ?             & y  \\
23	&	308.66538 & 60.13970 & 0.78	 &	0.22	&	0.20	&	0.20	&	0.47	&	0.49	&	-	    &	0.80	& 1.04      &	1.83	&	4.04	&	n	&	H	&  ?             & n  \\
24	&   308.73678 & 60.09243 & 0.54	 &	0.37	&	0.62	&	0.60	&	0.81	&	0.69	&	0.70	&	0.51	& 1.25      &	2.25	&	2.17	&	n	&	H	&  AGN (0.71)    & n  \\
26	&	308.65047 & 60.14428 & 0.45	 &	0.41	&	-	    &	0.03	&	0.54	&	0.06	&	-	    &	0.07	& 0.75      &	3.43	&	19.95	&	t	&	uc	&  NS (0.71)     & y  \\
27	&	308.65166 & 60.09962 & 0.38	 &	0.64	&	-	    &	-	    &	-	    &	0.48	&	-	    &	-	    & 0.92      &	1.87	&	1.71	&	n	&	H	&  ?             & y  \\
28	&	308.58915 & 60.17167 & 0.42	 &	0.80	&	-	    &	-	    &	0.00	&	0.00	&	-	    &	0.98	& 1.39      &	1.58	&	2.33	&	n	&	H	&  ?             & y  \\
30	&	308.60940 & 60.15417 & 0.52	 &	0.36	&	-	    &	-	    &	0.27	&	0.00	&	-	    &	0.41	& 0.99      &	1.97	&	1.93	&	n	&	uc	&  NS (0.79)     & n  \\
31	&	308.69803 & 60.16755 & 0.35	 &	0.39	&	0.55	&	0.58	&	0.45	&	0.16	&	0.20	&	0.31	&  0.79     &	2.44	&	3.63	&	n	&	L	&  ?             & y  \\
32	&	308.85647 & 60.16597 & 0.92	 &	0.39	&	1.07	&	0.87	&	1.11	&	-	    &  	0.95	&	0.92	&  1.84     &	3.33	&	2.88	&	v	&	L	&  ?             & n  \\
33	&	308.70844 & 60.15685 & 0.37	 &	0.54	&	0.55	&	0.49	&	0.70	&	0.63	&	0.59	&	0.79	& 1.11      &	1.31	&	2.14	&	n	&	H	&  ?             & n  \\
34	&	308.81763 & 60.17949 & 0.75	 &	3.34	&	2.58	&	0.03	&	0.26	&	-	    &	3.86	&	0.10	& 3.53      &	11.94	&	113.11	&	t	&	H	&  ?             & n  \\
35	&	308.70283 & 60.16166 & 0.29	 &	0.37	&	0.10	&	0.00	&	0.11	&	0.15	&	0.04	&	0.25	& 0.44      &	2.93	&	9.33	&	v	&	H	&  ?             & y  \\
37	&	308.62135 & 60.18088 & 0.30	 &	0.76	&	-	    &	-	    &	0.00	&	0.00	&	-	    &	0.37	& 0.94      &	3.09	&	2.49	&	v	&	H	&  AGN (0.73)    & y  \\
38	&	308.70733 & 60.16990 & 0.31	 &	0.36	&	0.35	&	0.53	&	0.38	&	0.19	&	0.16	&	0.21	& 0.68      &	2.38	&	3.22	&	n	&	H	&  AGN (0.80)    & y  \\
39	&	308.69163 & 60.18041 & 0.30	 &	0.28	&	0.18	&	0.04	&	0.20	&	0.00	&	-	    &	0.35	& 0.47      &	1.22	&	8.13	&	n	&	uc	&  ?             & y  \\
41	&	308.78382 & 60.18710 & 0.30	 &	0.54	&	0.49	&	0.31	&	0.51	&	0.17	&	0.29	&	0.54	&  0.79     &	1.25	&	3.25	&	n	&	L	&  ?             & y  \\
42	&	308.71022 & 60.17109 & 0.23	 &	0.24	&	0.32	&	0.17	&	0.00	&	0.09	&	0.28	&	0.10	& 0.48      &	0.73	&	1.90	&	n	&	uc	&  ?             & y  \\
44	&	308.65514 & 60.16507 & 0.19	 &	0.30	&	-	    &	0.20	&	0.06	&	0.03	&	-	    &	0.10	& 0.39      &	2.06	&	8.69	&	n	&	H	&  ?             & y  \\
46	&   308.70732 & 60.13506 & 0.18	 &	0.18	&	0.06	&	0.18	&	0.05	&	0.11	&	0.13	&	0.10	& 0.29      &	1.51	&	3.61	&	n	&	L	&  ?             & y  \\
47	&	308.67522 & 60.09152 & 0.19	 &	0.16	&	0.07	&	0.05	&	0.22	&	0.11	&	0.12	&	-	    & 0.30      &	1.34	&	4.88	&	n	&	H	&  ?             & y  \\
48	&	308.77711 & 60.17847 & 0.44	 &	0.42	&	0.39	&	0.62	&	0.93	&	0.24	&	0.47	&	0.41	& 1.03      &	3.17	&	3.87	&	v	&	H	&  AGN (0.71)    & n  \\
49	&	308.69853 & 60.15912 & 0.20	 &	2.59	&	0.11	&	0.34	&	2.56	&	1.67	&	0.15	&	0.27	& 1.94      &	9.91	&	24.26	&	t	&	L	&  ?             & n  \\
50	&	308.76963 & 60.15491 & 0.28	 &	0.19	&	0.12	&	0.16	&	0.21	&	0.02	&	0.33	&	0.36	& 0.45      &	0.93	&	2.94	&	n	&	L	&  ?             & n  \\
51	&	308.64715 & 60.12856 & 0.25	 &	0.28	&	-	    &	-	    &	0.12	&	0.05	&	-	    &	0.00	&  0.47     &	1.98	&	5.12	&	n	&	L	&  ?             & y  \\
52	&	308.86540 & 60.15625 & 0.19	 &	0.22	&	0.46	&	0.60	&	0.51	&	-	    &	0.00	&	0.95	&  0.79     &	2.14	&	5.00	&	n	&	H	&  AGN (0.70)    & n  \\
53	&	308.71176 & 60.20235 & 0.17	 &	0.26	&	0.30	&	0.30	&	0.04	&	0.04	&	-	    &	0.00	&  0.42     &	2.08	&	7.59	&	n	&	A	&  AGN (0.80)    & n  \\
55	&	308.75209 & 60.21467 & 0.45	 &	0.65	&	1.17	&	0.51	&	0.30	&	0.14	&	0.90	&	0.70	& 1.25      &	4.84	&	8.34	&	v	&	L	&  AGN (0.72)    & n  \\
56	&	308.63295 & 60.20214 & 0.13	 &	0.24	&	-	    &	0.11	&	0.00	&	0.17	&	-	    &	0.00	& 0.31      &	1.16	&	2.18	&	n	&	A	&  ?             & n  \\
57	&	308.72266 & 60.12216 & 0.15	 &	1.35	&	0.09	&	0.35	&	0.41	&	0.66	&	0.10	&	0.09	&  0.85     &	6.61	&	15.86	&	t	&	H	&  ?             & y  \\
58	&	308.64820 & 60.13576 & 0.14	 &	0.26	&	-	    &	-	    &	0.10	&	0.00	&	-	    &	0.00	& 0.33      &	1.27	&	2.47	&	n	&	L	&  ?             & y  \\
59	&	308.73220 & 60.15851 & 0.16	 &	0.23	&	0.48	&	0.69	&	0.72	&	0.66	&	0.90	&	0.56	&  0.98     &	4.53	&	5.76	&	v	&	A	&  ?             & n  \\
60	&	308.73751 & 60.20254 & 0.11	 &	0.19	&	0.27	&	0.14	&	0.05	&	0.05	&	0.13	&	0.01	& 0.27      &	1.36	&	31.50	&	t	&	L	&  ?             & n  \\
61	&	308.71478 & 60.21318 & 0.14	 &	0.42	&	0.28	&	0.33	&	0.09	&	0.00	&	0.00	&	0.74	& 0.52      &	1.99	&	8.56	&	n	&	H	&  AGN(0.76)     & y  \\
62	&	308.68428 & 60.12201 & 0.06	 &	2.91	&	0.71	&	0.58	&	0.60	&	0.30	&	0.12	&	0.20	& 1.47      &	11.23	&	45.19	&	t	&	H	&  ?             & y  \\
63	&	308.80701 & 60.08946 & -	 &	0.57	&	0.35	&	0.49	&	0.19	&	0.58	&	1.08	&	0.47	& 1.10      &	4.56	&	5.65	&	v	&	H	&  ?             & y  \\
64	&	308.80471 & 60.15198 & 0.12	 &	0.30	&	0.18	&	0.12	&	0.20	&	0.40	&	0.09	&	0.08	& 0.38      &	1.59	&	5.07	&	n	&	H	&  ?             & n  \\
65	&	308.75188 & 60.14975 & 0.15	 &	0.30	&	0.52	&	0.40	&	0.15	&	0.05	&	0.40	&	0.50	& 0.58      &	3.22	&	9.54	&	v	&	L	&  ?             & y  \\
66	&	308.83826 & 60.16433 & 0.22	 &	0.18	&	0.46	&	0.36	&	0.12	&	-	    &	0.13	&	0.18	& 0.49      &	2.22	&	3.93	&	n	&	L	&  ?             & y  \\
67	&	308.85563 & 60.12154 & -	 &	0.03	&	5.89	&	4.78	&	5.07	&	0.18	&	0.19	&	0.06	& 4.87      &	14.46	&	215.62	&	t	&	H	&  ?             & n  \\
69	&	308.88455 & 60.17167 & 0.18	 &	-	    &	0.00	&	0.66	&	0.68	&	-	    &	0.92	&	0.08	& 1.05      &	3.81	&	10.97	&	t	&	H	&  ?             & n  \\
%70	&	308.87776 & 60.12703 & -	 &	0.12	&	0.41	&	0.43	&	0.53	&	-	    &	0.27	&	0.16	& 0.64      &	2.64	&	4.27	&	v	&	L	&  LM-Star (0.97)& n  \\
71	&	308.79550 & 60.14888 & 0.31	 &	0.25	&	0.33	&	0.39	&	0.33	&	0.10	&	0.18	&	0.00	& 0.58      &	2.10	&	3.87	&	n	&	L	&  ?             & n  \\
72	&	308.82243 & 60.13729 & 0.09	 &	0.09	&	0.24	&	0.32	&	0.09	&	0.13	&	0.31	&	0.14	& 0.35      &	1.70	&	3.40	&	n	&	H	&  AGN ( 0.95)   & n  \\
73	&	308.77290 & 60.12837 & 0.09	 &	0.08	&	0.32	&	0.18	&	0.22	&	0.16	&	0.14	&	0.41	& 0.34      &	1.23	&	5.23	&	n	&	H	&  AGN ( 0.83)   & y  \\
74	&	308.73449 & 60.15202 & 0.14	 &	0.04	&	0.28	&	0.26	&	0.00	&	0.00	&	0.07	&	0.00	& 0.28      &	2.20	&	7.26	&	n	&	L	&  ?             & y  \\
75	&	308.75756 & 60.09275 & 0.04	 &	0.04	&	0.24	&	0.18	&	0.06	&	0.34	&	0.02	&	0.00	& 0.22      &	2.31	&	15.93	&	t	&	H	&  AGN (0.73)    & n  \\
76	&	308.79248 & 60.16678 & 0.23	 &	0.14	&	0.22	&	0.29	&	0.17	&	0.20	&	0.13	&	0.16	& 0.42      &	1.29	&	2.22	&	n	&	uc	&  ?             & n  \\
77	&	308.63144 & 60.18233 & 0.07	 &	0.08	&	0.00	&	-	    &	0.90	&	0.25	&	-	    &	0.04	& 0.51      &	3.62	&	20.36	&	t	&	H	&  ?             & n  \\
78	&	308.71287 & 60.17876 & 0.00	 &	0.04	&	0.06	&	0.00	&	0.00	&	4.21	&	18.48	&	9.37	& 11.81     &	27.26	&	491.23	&	t	&	A	&  AGN (0.70)    & y  \\
79	&	308.71039	&	60.12337	&	0.09	&	1.48	&	1.77	&	1.78	&	1.44	&	0.11	&	0.12	&	0.62	&	0.83	&	7.00	&	20.60	&	t	&	H	&		&	y	\\
80	&	308.65793	&	60.07627	&	0.66	&	0.99	&	0.88	&	-	&	-	&	2.15	&	1.29	&	-	&	1.07	&	4.86	&	3.25	&	v	&	H	&		&	y	\\
81	&	308.71017	&	60.15262	&	0.34	&	0.13	&	0.59	&	0.29	&	0.45	&	4.67	&	0.34	&	0.00	&	0.70	&	10.45	&	37.04	&	t	&	A	&		&	n	\\
82	&	308.68400	&	60.19336	&	0.04	&	0.00	&	0.32	&	0.06	&	0.12	&	0.00	&	-	&	9.08	&	0.41	&	9.13	&	209.81	&	t	&	H	&		&	y	\\
83	&	308.69345	&	60.22438	&	0.79	&	0.36	&	1.12	&	0.52	&	0.60	&	0.26	&	-	&	0.62	&	0.68	&	3.18	&	3.13	&	v	&	H	&	AGN (0.84)	&	n	\\
84	&	308.70754	&	60.17230	&	0.48	&	0.29	&	1.24	&	0.60	&	0.36	&	0.26	&	0.19	&	0.21	&	0.48	&	4.46	&	6.34	&	v	&	L	&	AGN (0.76)	&	y	\\
85	&	308.77013	&	60.07164	&	-	&	0.06	&	0.10	&	0.37	&	0.27	&	0.19	&	0.19	&	0.00	&	0.17	&	1.54	&	5.91	&	n	&	H	&	AGN (0.79)	&	n	\\
86	&	308.69627	&	60.14740	&	0.35	&	0.00	&	0.18	&	0.23	&	0.00	&	0.04	&	0.17	&	0.52	&	0.19	&	1.32	&	3.03	&	n	&	H	&	AGN (0.76)	&	n	\\
87	&	308.80209	&	60.14374	&	0.18	&	0.15	&	0.29	&	0.34	&	0.24	&	0.33	&	0.39	&	0.21	&	0.26	&	1.52	&	2.64	&	n	&	L	&		&	y	\\
88	&	308.79176	&	60.15207	&	0.34	&	0.08	&	0.40	&	0.17	&	0.27	&	0.22	&	0.39	&	0.21	&	0.28	&	1.74	&	4.71	&	n	&	H	&	AGN (0.74)	&	n	\\
89	&	308.83348	&	60.15934	&	0.37	&	0.02	&	0.43	&	0.23	&	0.24	&	0.15	&	0.17	&	0.31	&	0.25	&	2.27	&	20.18	&	t	&	uc	&		&	y	\\
90	&	308.68008	&	60.11458	&	0.25	&	0.04	&	0.32	&	0.29	&	0.15	&	0.19	&	0.05	&	0.21	&	0.19	&	1.61	&	7.52	&	n	&	H	&	AGN (0.78)	&	y	\\
91	&	308.65528	&	60.17194	&	0.24	&	0.08	&	-	&	0.17	&	0.15	&	0.15	&	-	&	0.31	&	0.19	&	0.77	&	3.65	&	n	&	L	&		&	y	\\
92	&	308.72304	&	60.14302	&	0.31	&	0.21	&	0.37	&	0.23	&	0.15	&	0.00	&	0.39	&	0.41	&	0.27	&	1.31	&	2.73	&	n	&	H	&		&	n	\\
%93	&	308.73859	&	60.12605	&	0.25	&	0.25	&	0.34	&	0.32	&	0.15	&	0.11	&	0.24	&	0.21	&	0.25	&	1.33	&	3.09	&	n	&	uc	&		&	n	\\
94	&	308.71326	&	60.14384	&	0.14	&	0.27	&	0.34	&	0.26	&	0.39	&	0.00	&	0.15	&	0.31	&	0.22	&	2.05	&	2.78	&	n	&	L	&		&	n	\\
95	&	308.75477	&	60.15648	&	0.18	&	0.23	&	0.37	&	0.17	&	0.39	&	0.11	&	0.46	&	0.41	&	0.27	&	2.19	&	4.15	&	n	&	L	&		&	y	\\
96	&	308.66956	&	60.14081	&	0.29	&	0.08	&	0.32	&	0.09	&	0.33	&	0.33	&	-	&	0.00	&	0.25	&	1.35	&	3.94	&	n	&	L	&	AGN (0.79)	&	y	\\
97	&	308.71663	&	60.18268	&	0.10	&	0.27	&	0.24	&	0.20	&	0.00	&	0.11	&	0.44	&	0.52	&	0.20	&	1.58	&	5.31	&	n	&	H	&		&	y	\\
98	&	308.74779	&	60.14244	&	0.18	&	0.27	&	0.45	&	0.29	&	0.12	&	0.37	&	0.19	&	0.10	&	0.25	&	1.25	&	4.37	&	n	&	L	&	AGN (0.88)	&	y	\\
99	&	308.70594	&	60.22847	&	0.46	&	0.15	&	0.10	&	0.14	&	0.06	&	0.26	&	-	&	0.72	&	0.28	&	1.83	&	11.97	&	t	&	H	&	AGN (0.82)	&	y	\\
100	&	308.79883	&	60.15039	&	0.41	&	0.29	&	0.70	&	0.49	&	0.39	&	0.48	&	0.56	&	0.21	&	0.45	&	1.59	&	3.36	&	n	&	L	&		&	y	\\
%101	&	308.69005	&	60.15612	&	0.27	&	0.17	&	0.40	&	0.37	&	0.39	&	0.04	&	0.07	&	0.52	&	0.26	&	1.50	&	13.92	&	t	&	L	&		&	n	\\
102	&	308.83611	&	60.12799	&	0.16	&	0.00	&	0.00	&	0.17	&	0.15	&	0.37	&	0.37	&	0.72	&	0.18	&	2.01	&	4.78	&	n	&	H	&		&	n	\\
103	&	308.73684	&	60.13649	&	0.18	&	0.13	&	0.72	&	0.29	&	0.18	&	0.15	&	0.24	&	0.41	&	0.26	&	2.91	&	5.72	&	n	&	uc	&		&	y	\\
104	&	308.89746	&	60.17340	&	0.04	&	-	&	0.40	&	0.11	&	0.00	&	-	&	0.44	&	1.14	&	0.18	&	2.72	&	26.22	&	t	&	H	&		&	n	\\
105	&	308.66918	&	60.16301	&	0.10	&	0.04	&	0.21	&	0.40	&	0.42	&	0.15	&	-	&	0.10	&	0.19	&	2.11	&	10.02	&	t	&	H	&		&	y	\\
106	&	308.88313	&	60.15050	&	0.07	&	-	&	0.07	&	0.34	&	0.03	&	-	&	0.10	&	0.21	&	0.11	&	1.92	&	11.41	&	t	&	L	&	LM-Star (0.79)	&	n	\\
107	&	308.81158	&	60.08706	&	-	&	0.00	&	0.07	&	0.32	&	0.00	&	0.11	&	0.07	&	0.00	&	0.08	&	1.96	&	4.32	&	n	&	H	&		&	y	\\
108	&	308.76382	&	60.14506	&	0.10	&	0.04	&	0.29	&	0.20	&	0.06	&	0.15	&	0.22	&	0.00	&	0.14	&	1.16	&	6.87	&	n	&	H	&		&	n	\\
109	&	308.72324	&	60.10747	&	0.14	&	0.00	&	0.21	&	0.37	&	0.12	&	0.11	&	0.00	&	0.21	&	0.13	&	2.30	&	3.35	&	n	&	H	&	AGN (0.74)	&	n	\\
110	&	308.77771	&	60.16923	&	0.04	&	0.15	&	0.18	&	0.11	&	0.03	&	0.07	&	0.32	&	0.41	&	0.13	&	1.23	&	13.67	&	t	&	L	&	AGN (0.80)	&	y	\\
111	&	308.73519	&	60.07949	&	-	&	0.00	&	0.00	&	0.23	&	0.15	&	0.33	&	0.07	&	0.00	&	0.10	&	1.93	&	4.57	&	n	&	H	&		&	n	\\
112	&	308.66698	&	60.07494	&	0.08	&	0.02	&	0.24	&	-	&	-	&	0.04	&	0.10	&	-	&	0.09	&	1.32	&	11.19	&	t	&	H	&		&	y	\\
113	&	308.82881	&	60.12812	&	0.05	&	0.00	&	0.40	&	0.11	&	0.03	&	0.26	&	0.27	&	0.00	&	0.13	&	2.14	&	13.19	&	t	&	H	&		&	n	\\
\enddata 
\end{deluxetable*} 
\end{longrotatetable}

%%%

%\begin{figure*}
% \begin{center}
% \includegraphics[width=110mm]{Figure4a.png}
% \includegraphics[width=110mm]{Figure4b.png}
% \end{center}
% \caption{The differantial XLFs for HMXBs (mass $\geq$ 8M$_{\odot}$ according to the CMD), IMXBs (1M$_{\odot}$ $\leq$ mass $\leq$ 8M$_{\odot}$ ) and LMXBs ( mass $\leq$ 1M$_{\odot}$) which are classified based on their optical data. \textbf{The red dashed lines indicate the best-fit single power-law models obtained from an unbinned maximum likelihood estimation (MLE). The goodness of fit was evaluated using a Monte Carlo Kolmogorov–Smirnov (K–S) test.}}
% \label{xlf2}
%\end{figure*}

\begin{figure*}
 \begin{center}
  \includegraphics[width=150mm]{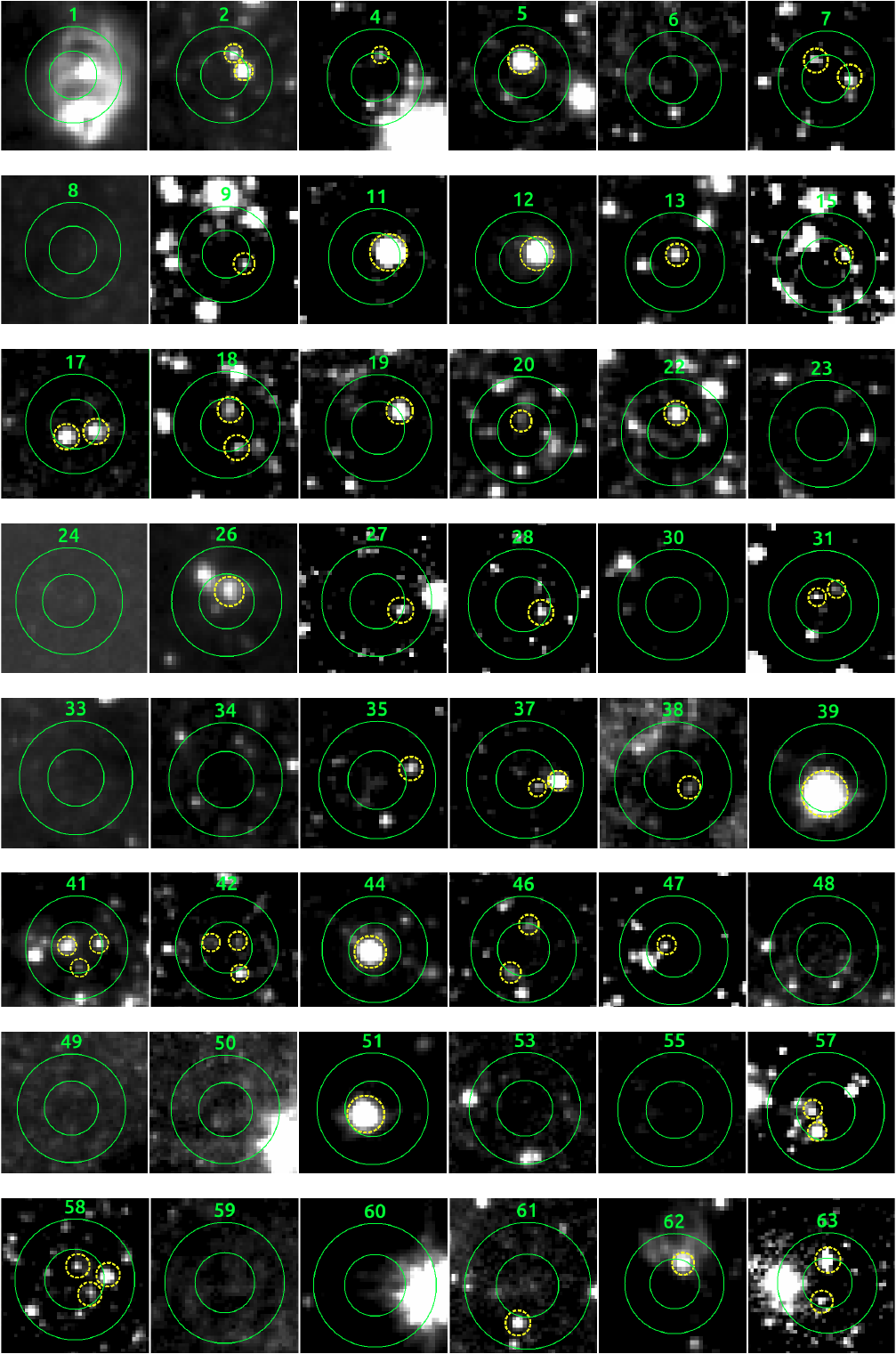}
   \end{center}
 \caption{The {\it HST}/ACS F606W filter images of the corrected positions of X-ray sources. Green circles show the astrometric error radii with 1$\sigma$ and 2$\sigma$ uncertainity and yellow dashed circles represent possible optical counterparts.}
 \label{opt_counterparts}
\end{figure*}

\begin{figure*}
    \ContinuedFloat
    \centering
    \includegraphics[width=150mm]{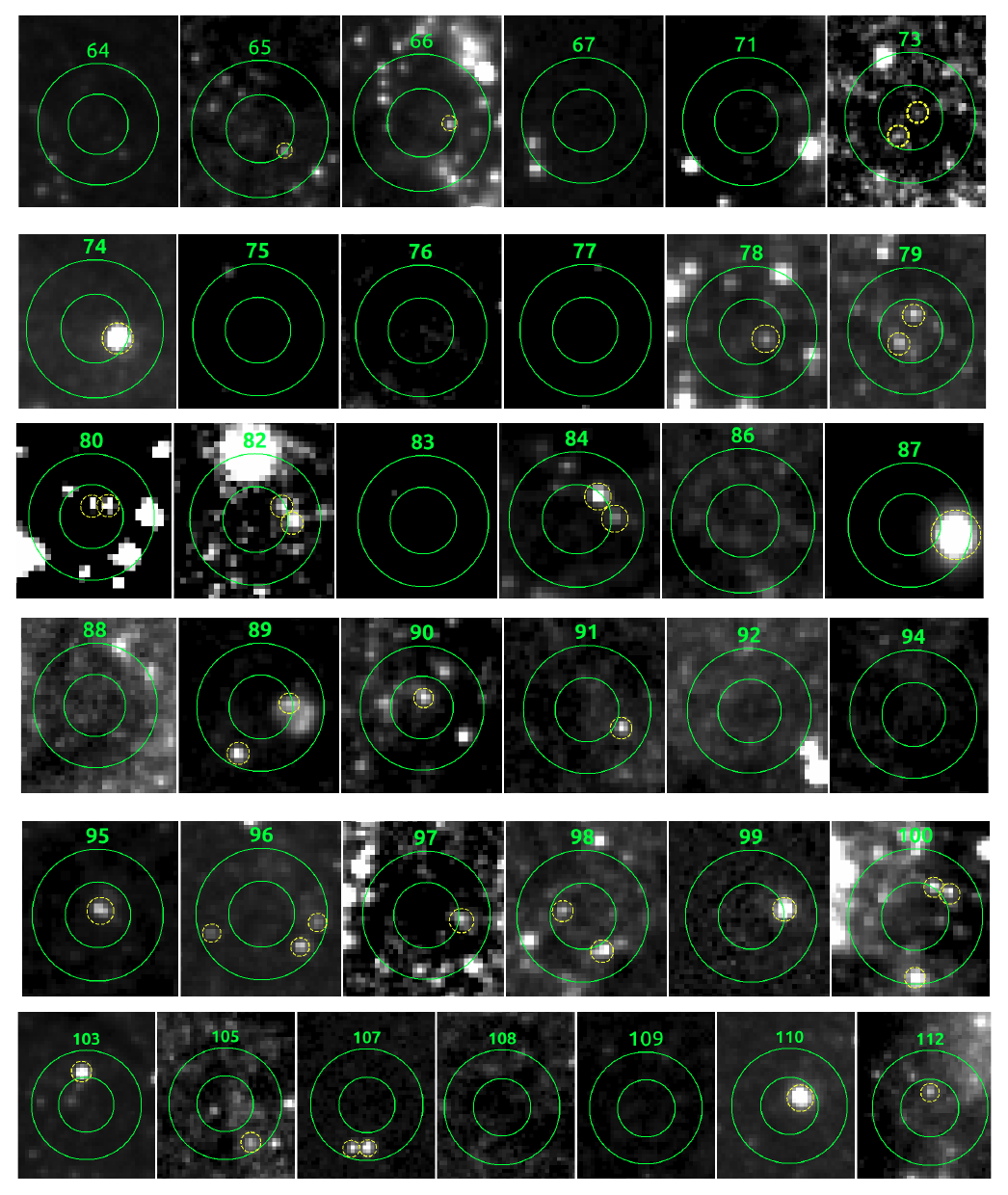}
    \caption{Continued}
 \label{opt_counterparts}
\end{figure*}

\begin{figure*}
\begin{center}
\includegraphics[angle=0,scale=0.44]{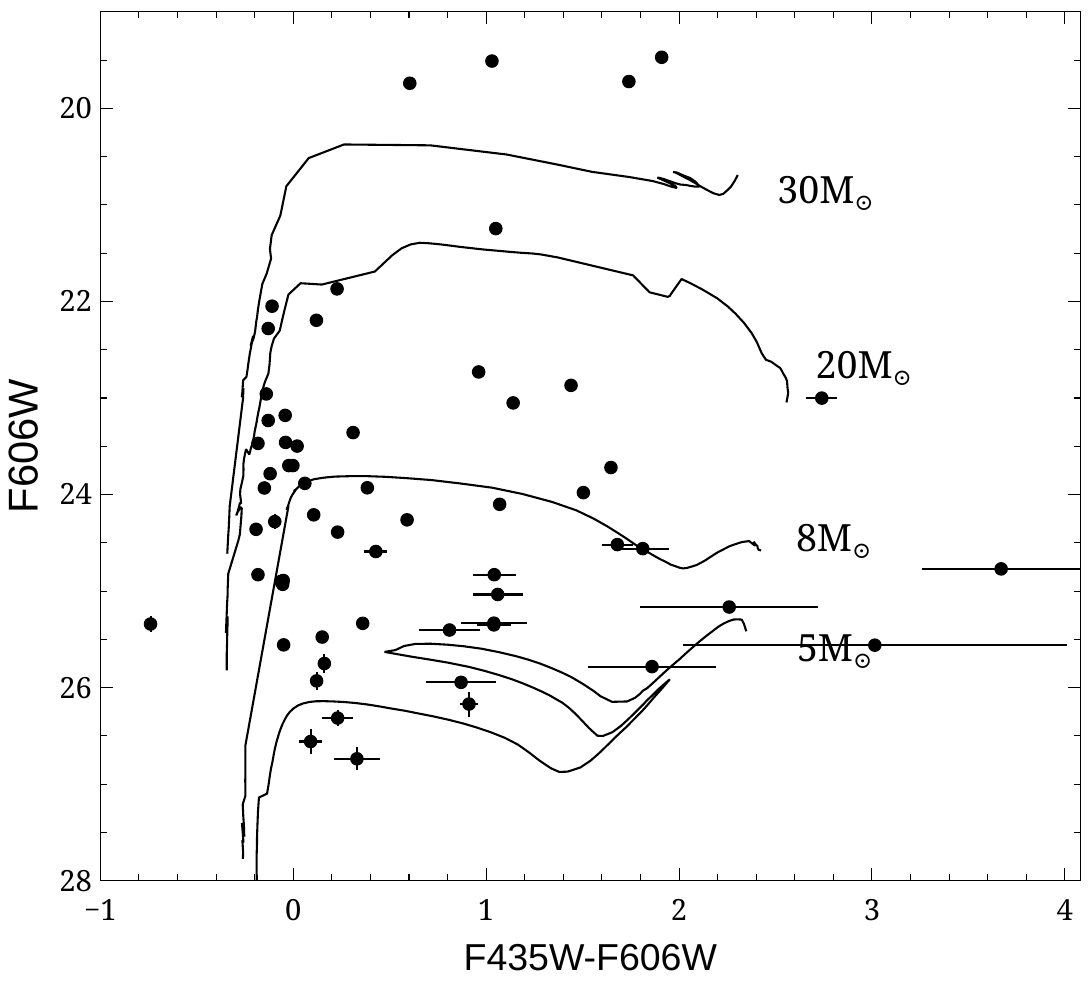}
\includegraphics[angle=0,scale=0.39]{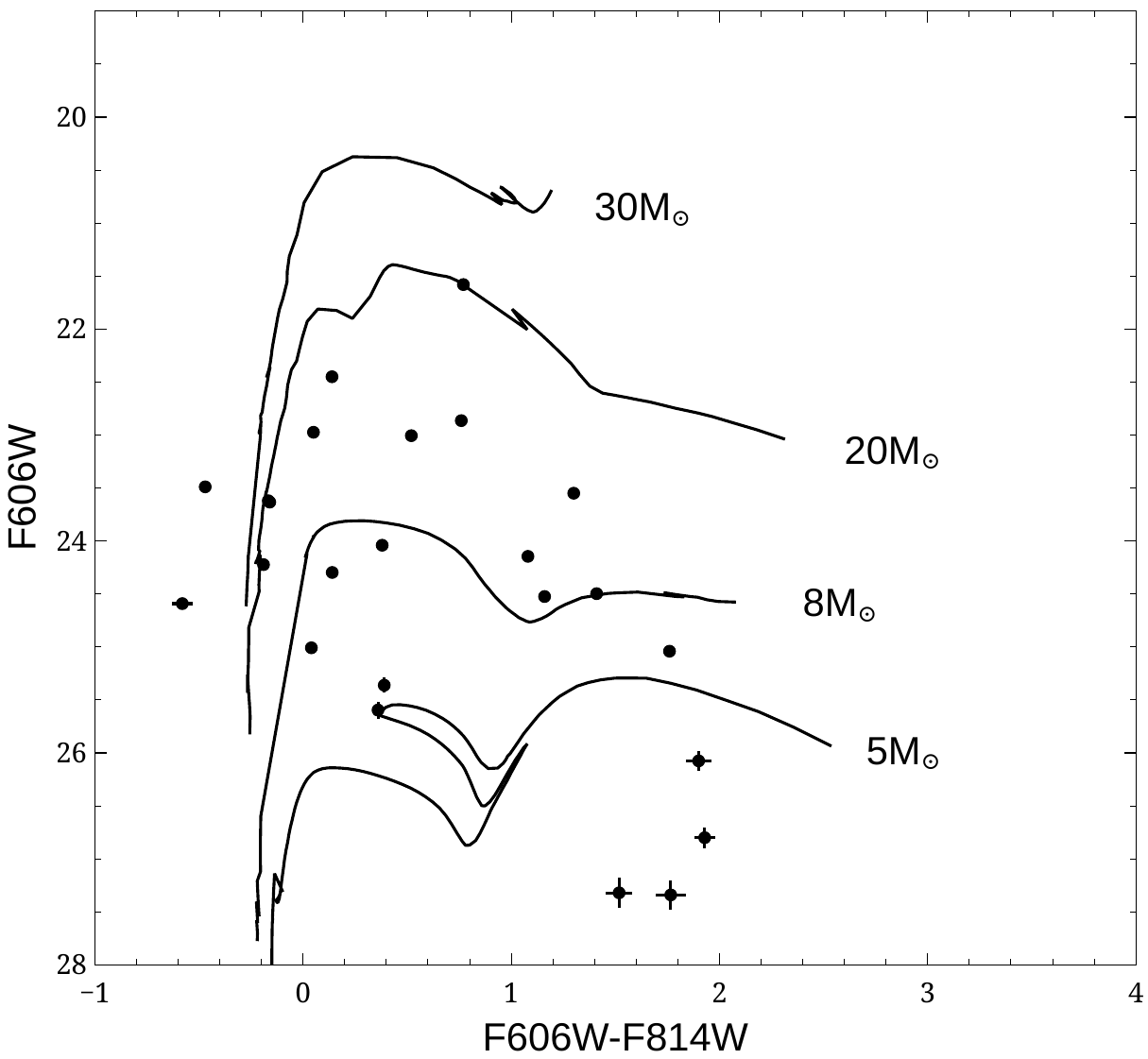}
\caption{The {\it HST}/ACS color-magnitude diagrams for possible optical counterparts of XRB candidates in NGC 6946. The left panel shows sources detected in the F435W and F606W filters, and the right panel shows sources detected in the F606W and F814W filters. Because not all counterparts are detected in all filters (Table \ref{opt_par}), the two CMDs represent different source samples. The black dots are represent optical candidates and the PARSEC isochrones for different masses are overplotted as black solid lines.}
\label{cmd}
\end{center}
\end{figure*}

\begin{figure}
 \begin{center}
  \includegraphics[width=87mm]{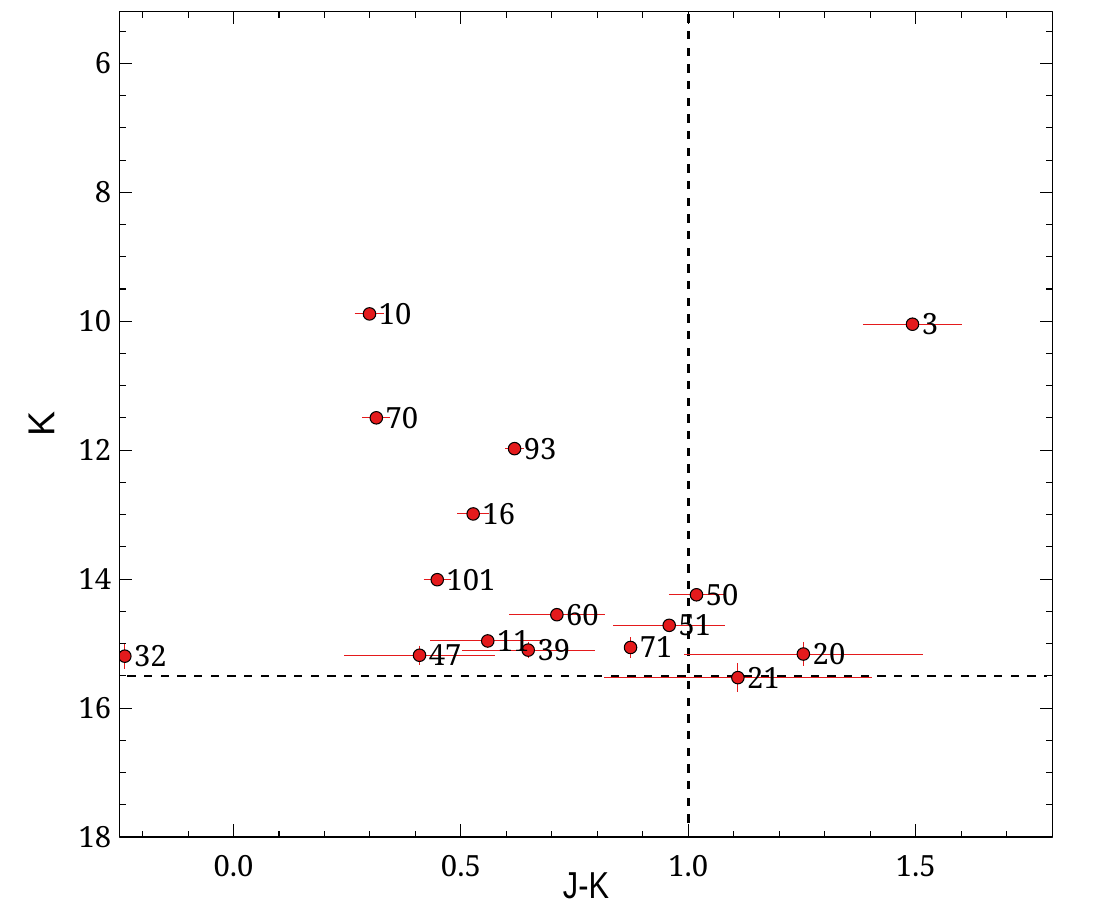}
 \end{center}
 \caption{The Color-magnitude diagram of infrared counterparts of 9 XRB candidates, two known SNRs (\#3 and \#21) and five foreground stars (\#10, \#16, \#70, \#93 and \#101). The sources are marked with the IDs used in Table \ref{x-ray_par}. The dashed lines indicate the J/K limits to define the AGN region (top right segment).}
 \label{j-k}
\end{figure}

\begin{table*}
	\centering
 	\caption{X-ray spectral best-fit parameters of XRB candidates which have significant statistics in C1 and C2 observations calculated with {\scshape tbabs$\times$tbabs$\times$po} and {\scshape tbabs$\times$tbabs$\times$diskbb} models}
  \label{best-fit}
	\begin{tabular}{cccccccc} % four columns, alignment for each
		\hline  \hline     
 SrcID.  & Label  & Model & $N_{\mathrm{H}}$  & $\Gamma$ / $kT_{\mathrm{in}}$  &  ${\Delta \chi}^{2}$ (dof) & $L_{\mathrm{X}}$ \\
	&  &  & ($10^{22}$ cm$^{-2}$) & (keV) &  & (10$^{39}$ erg~cm$^{-2}$)  \\

\hline
1	&	C1	&	{\scshape po} &	$0.014_{-0.005}^{+0.005}$	&	$2.536_{-0.032}^{+0.033	}$	&	1.77 (167)	&	$12.82_{-0.01}^{+0.008}$	\\
	&		&		&								&								&					&							\\
 2	&	C1	&	{\scshape po}&	$0.384_{-0.011}^{+0.011}$	&	$5.078_{-0.100}^{+0.101	}$	&	1.97 (63)	&	$44.93_{-0.02}^{+0.016}$	\\
   \vspace{0.3cm}	
	&	C2	&	{\scshape po} &		0.000					&	$4.679_{-0.256}^{+0.300	}$	&	2.21 (12)	&	$9.05_{-0.04}^{+0.038}$	\\
  \vspace{0.1cm}	
3	&	C1	&	{\scshape po}	&	$0.209_{-0.027}^{+0.029}$	&	$2.372_{-0.067}^{+0.073	}$	&	1.32 (50)	&	$2.73_{-0.02}^{+0.020}$	\\
\vspace{0.3cm}	
	&	C2	&	{\scshape po}	&	$0.139_{-0.078}^{+0.090}$	&	$2.189_{-0.094}^{+0.105	}$	&	1.51 (19)	&	$2.09_{-0.04}^{+0.033}$	\\
    \vspace{0.1cm}	
4	&	C1	&	{\scshape po}	&	$0.106_{-0.028}^{+0.031}$	&	$2.055_{-0.079}^{+0.086	}$	&	1.16 (36)	&	$1.41_{-0.03}^{+0.024}$	\\
	&		&		&								&								&					&							\\
5	&	C1	&	{\scshape po}	&	$0.315_{-0.015}^{+0.017}$	&	$4.684_{-0.150}^{+0.150	}$	&	1.74 (33)	&	$11.37_{-0.03}^{+0.024}$	\\
	&		&		&						&								&					&							\\
6	&	C1	&	{\scshape diskbb}	&	$0.029_{-0.029}^{+0.085}$	&	$2.110_{-0.058}^{+0.055	}$	&	1.00 (16)	& $0.67_{-0.04}^{+0.035}$	\\
\vspace{0.3cm}	
	&	C2	&	{\scshape diskbb}	&	$0.258_{-0.168}^{+0.206}$	&	$1.742_{-0.113}^{+0.130	}$	&	1.16 (8)	& $0.78_{-0.05}^{+0.047}$	\\
    \vspace{0.1cm}	
7	&	C1	&	{\scshape diskbb}	&		0.000					&	$1.327_{-0.035}^{+0.033	}$	&	1.20 (15)	& $0.49_{-0.04}^{+0.037}$	\\
\vspace{0.3cm}	
	&	C2	&	{\scshape po}	&	$0.139_{-0.138}^{+0.172}$	&	$1.520_{-0.113}^{+0.131	}$	&	0.95 (8)	&	$ 0.88_{-0.05}^{+0.046}$	\\
    \vspace{0.1cm}	
8	&	C1	&	{\scshape diskbb}	&	$0.050_{-0.048}^{+0.056}$	&	$0.845_{-0.020}^{+0.019	}$	&	0.94 (14)	& $0.42_{-0.04}^{+0.037}$	\\
\vspace{0.3cm}	
	&	C2	&	{\scshape po}	&	$0.174_{-0.124}^{+0.148}$	&	$1.971_{-0.145}^{+0.168	}$	&	1.32 (8)	&	$1.06_{-0.05}^{+0.047}$	\\
    \vspace{0.1cm}	
9	&	C1	&	{\scshape diskbb}	&	$0.533_{-0.074}^{+0.085}$	&	$0.759_{-0.014}^{+0.014	}$	&	0.62 (19)	& $	0.93_{-0.04}^{+0.033}$	\\
\vspace{0.3cm}	
	&	C2	&	{\scshape diskbb}	&	$0.445_{-0.109}^{+0.126}$	&	$1.139_{-0.026}^{+0.024	}$	&	0.80 (16)	& $	1.35_{-0.04}^{+0.035}$	\\
    \vspace{0.1cm}	
11	&	C1	&	{\scshape po}&		0.000						&	$1.542_{-0.157}^{+0.187	}$	&	0.78 (8)	& $	0.41_{-0.05}^{+0.046}$	\\
	&		&		&								&								&					&							\\
12	&	C1	&	{\scshape po}	&	$0.095_{-0.080}^{+0.098}$	&	$1.942_{-0.179}^{+0.217	}$	&	1.43 (6)	&	$0.35_{-0.06}^{+0.051}$	\\
\vspace{0.3cm}	
	&	C2	&	{\scshape diskbb}	&	$0.141_{-0.066}^{+0.292}$	&	$1.234_{-0.055}^{+0.050	}$	&	1.65 (3)	&	$0.45_{-0.07}^{+0.062}$	\\
    \vspace{0.1cm}	
14	&	C1	&	{\scshape po}	&	$0.047_{-0.047}^{+0.109}$	&	$1.606_{-0.16	}^{+0.194	}$	&	0.81 (6)	&	$0.35_{-0.06}^{+0.052}$	\\
\vspace{0.3cm}	
	&	C2	&	{\scshape diskbb}	&		0.000				&	$1.001_{-0.059}^{+0.051}$	&	1.11 (2)	&	$0.17_{-0.11}^{+0.086}$	\\
    \vspace{0.1cm}	
15	&	C1	&	po	&		0.000					&	$1.287_{-0.154}^{+0.190	}$	&	1.07 (5)	&	$0.30_{-0.06}^{+0.055}$	\\
	&		&		&								&								&					&							\\
17	&	C1	&	{\scshape diskbb}	&	$0.282_{-0.215}^{+0.296}$	&	$1.574_{-0.098}^{+0.087}$	&	1.98 (1)	&	$0.18_{-0.09}^{+0.076}$	\\
&		&		&								&								&					&							\\

\hline
	\end{tabular}
\end{table*}

\LTcapwidth=\textwidth
\clearpage
\begin{longtable*}{c c c c c c c }
\caption{De-reddened magnitude values and colors of optical candidates of XRB candidates in NGC 6946 obtained with {\it HST}/ACS data} \label{opt_par} \\
\toprule
\hline  \hline   
 SrcID & F435W & F606W & F814W & M$_{V}$ & (B-V)$_{0}$ &  (V-I)$_{0}$  \\
 (1) & (2) & (3) & (4) & (5) & (6) & (7)  \\
\midrule
\endfirsthead

\multicolumn{7}{c}{{\bfseries Table \thetable{} -- continued}} \\
\toprule
SrcID & F435W & F606W & F814W & M$_{V}$ & (B-V)$_{0}$ &  (V-I)$_{0}$  \\
 (1) & (2) & (3) & (4) & (5) & (6) & (7)  \\
\midrule
\endhead

\endfoot

\bottomrule
\endlastfoot
2	&	22.32	$\pm$	0.01	&	22.20	$\pm$	0.01	&				&	-7.26	&	-0.15	&		\\ %B5	
	&	24.20	$\pm$	0.03	&	23.06	$\pm$	0.01	&				&	-6.41	&	0.87	&		\\%K1	
4	&	26.10	$\pm$	0.17	&	25.04	$\pm$	0.04	&				&	-4.42	&	0.79	&		\\%G8	
5	&	21.94	$\pm$	0.01	&	22.05	$\pm$	0.01	&				&	-7.41	&	-0.38	&		\\%O	
7	&				&	24.60	$\pm$	0.03	&	25.18	$\pm$	0.09	&	-4.87	&		&	-0.78	\\		
	&				&	24.30	$\pm$	0.03	&	24.17	$\pm$	0.04	&	-5.16	&		&	-0.06	\\%A0	
9	&	25.51	$\pm$	0.06	&	25.56	$\pm$	0.05	&				&	-3.90	&	-0.32	&		\\%O	
11	&				&	21.58	$\pm$	0.01	&	20.81	$\pm$	0.01	&	-7.88	&		&	0.57	\\%G2	
12	&	22.30	$\pm$	0.01	&	21.25	$\pm$	0.01	&				&	-8.21	&	0.78	&		\\%K0	
13	&	23.79	$\pm$	0.02	&	23.94	$\pm$	0.02	&				&	-5.52	&	-0.42	&		\\%O	
15	&	25.63	$\pm$	0.08	&	25.48	$\pm$	0.05	&				&	-3.98	&	-0.12	&		\\%B7	
	&	26.66	$\pm$	0.19	&	26.56	$\pm$	0.13	&				&	-2.90	&	-0.18	&		\\%B4	
17	&	25.17	$\pm$	0.04	&	24.11	$\pm$	0.01	&				&	-5.36	&	0.80	&		\\%K0	
	&	26.21	$\pm$	0.10	&	24.52	$\pm$	0.02	&				&	-4.94	&	1.41	&		\\	%M0	
18	&	26.38	$\pm$	0.22	&	25.34	$\pm$	0.05	&				&	-4.12	&	0.77	&		\\%G8	
	&	24.61	$\pm$	0.10	&	25.35	$\pm$	0.08	&				&	-4.12	&	-1.01	&		\\	
19	&				&	23.49	$\pm$	0.02	&	23.96	$\pm$	0.03	&	-5.97	&		&	-0.67	\\	
20	&	22.83	$\pm$	0.02	&	22.96	$\pm$	0.01	&				&	-6.50	&	-0.14	&		\\%B6	
22	&	22.16	$\pm$	0.01	&	22.28	$\pm$	0.01	&				&	-7.18	&	-0.40	&		\\%O	
26	&	24.31	$\pm$	0.03	&	22.87	$\pm$	0.01	&				&	-6.59	&	1.17	&		\\%K5	
27	&				&	26.08	$\pm$	0.09	&	24.18	$\pm$	0.02	&	-3.38	&		&	1.70	\\%M0	
28	&	26.22	$\pm$	0.22	&	25.41	$\pm$	0.05	&				&	-4.05	&	0.54	&		\\%F8	
31	&				&	24.50	$\pm$	0.03	&	23.09	$\pm$	0.02	&	-4.96	&		&	1.21	\\%K5	
	&				&	25.01	$\pm$	0.06	&	24.97	$\pm$	0.09	&	-4.45	&		&	-0.16	\\%B8	
35	&				&	24.15	$\pm$	0.02	&	23.07	$\pm$	0.02	&	-5.31	&		&	0.88	\\%K2	
37	&	24.63	$\pm$	0.04	&	24.39	$\pm$	0.03	&				&	-5.07	&	-0.04	&		\\%B9	
	&	26.82	$\pm$	0.25	&	25.95	$\pm$	0.07	&				&	-3.51	&	0.60	&		\\%G0	
38	&				&	25.04	$\pm$	0.05	&	23.29	$\pm$	0.02	&	-4.42	&		&	1.56	\\%K7	
39	&				&	18.88	$\pm$	0.01	&	17.66	$\pm$	0.01	&	-10.58	&		&	1.02	\\%K4	
41	&	25.75	$\pm$	0.09	&	23.01	$\pm$	0.01	&				&	-6.46	&	2.47	&		\\	
	&	23.11	$\pm$	0.01	&	23.24	$\pm$	0.01	&				&	-6.22   &	-0.40	&		\\%O	
	&	29.60	$\pm$	3.03	&	24.60	$\pm$	0.02	&				&	-4.86	&	4.73	&		\\	
42	&				&	25.37	$\pm$	0.07	&	24.97	$\pm$	0.08	&	-4.10	&		&	0.19	\\%A8	
	&				&	25.60	$\pm$	0.08	&	25.24	$\pm$	0.10	&	-3.86	&		&	0.16	\\%A7	
44	&	20.55	$\pm$	0.01	&	19.51	$\pm$	0.01	&				&	-9.95	&	0.76	&		\\%G8	
46	&	25.70	$\pm$	0.07	&	25.34	$\pm$	0.04	&				&	-4.12	&	0.09	&		\\%A3	
	&	27.65	$\pm$	0.39	&	25.79	$\pm$	0.06	&				&	-3.67	&	1.59	&		\\%M4	
47	&				&	23.64	$\pm$	0.02	&	23.80	$\pm$	0.02	&	-5.82	&		&	-0.36	\\%O	
51	&	21.47	$\pm$	0.01	&	19.73	$\pm$	0.01	&				&	-9.73	&	1.47	&		\\%M1	
%53	&				&	23.01	$\pm$	0.01	&	22.49	$\pm$	0.01	&	-6.45	&		&	0.32	\\	%F1	
	%&				&	24.53	$\pm$	0.03	&	23.37	$\pm$	0.02	&	-4.93	&		&	0.96	\\%K3	
%	&				&	24.04	$\pm$	0.02	&	23.67	$\pm$	0.02	&	-5.42	&		&	0.18	\\	%A8	
57	&				&	22.87	$\pm$	0.01	&	22.11	$\pm$	0.01	&	-6.59	&		&	0.56	\\%G2	
	&				&	23.55	$\pm$	0.02	&	22.26	$\pm$	0.01	&	-5.91	&		&	1.10	\\	%K4	
58	&	28.44	$\pm$	1.14	&	24.77	$\pm$	0.03	&				&	-4.69	&	3.40	&		\\		
	&	27.43	$\pm$	0.50	&	25.17	$\pm$	0.05	&				&	-4.29	&	1.99	&		\\	
	&	23.68	$\pm$	0.02	&	23.36	$\pm$	0.01	&				&	-6.10	&	0.04	&		\\	%A2
61	&	23.67	$\pm$	0.01	&	23.79	$\pm$	0.02	&				&	-5.67	&	-0.39	&		\\	%O	
62	&				&	22.98	$\pm$	0.01	&	22.93	$\pm$	0.01	&	-6.48	&		&	-0.15	\\%B8	
63	&	26.21	$\pm$	0.11	&	26.39	$\pm$	0.13	&				&	-3.07	&	-0.46	&		\\	
65	&	24.86	$\pm$	0.04	&	24.27	$\pm$	0.02	&				&	-5.19	&	0.32	&		\\%F0	
66	&	23.95	$\pm$	0.02	&	23.89	$\pm$	0.01	&				&	-5.57	&	-0.21	&		\\%B3	
73	&	26.55	$\pm$	0.16	&	26.32	$\pm$	0.08	&				&	-3.14	&	-0.04	&		\\%B9	
	&	27.08	$\pm$	0.24	&	26.74	$\pm$	0.12	&				&	-2.72	&	0.06	&		\\%A2	
74	&	21.39	$\pm$	0.01	&	19.48	$\pm$	0.01	&				&	-9.99	&	1.64	&		\\	
78	&				&	24.23	$\pm$	0.02	&	24.42	$\pm$	0.04	&	-5.23	&		&	-0.39	\\%O	
79	&	24.65	$\pm$	0.04	&	24.83	$\pm$	0.03	&				&	-4.63	&	-0.46	&		\\
	&	24.88	$\pm$	0.05	&	24.93	$\pm$	0.03	&				&	-4.53	&	-0.33	&		\\
80	&				&	27.34	$\pm$	0.14	&	25.57	$\pm$	0.06	&	-2.12	&		&	1.57	\\
	&				&	27.32	$\pm$	0.14	&	25.80	$\pm$	0.08	&	-2.14	&		&	1.32	\\
82	&	27.08	$\pm$	0.18	&	26.17	$\pm$	0.13	&				&	-3.29	&	0.64	&		\\
	&	26.05	$\pm$	0.09	&	25.93	$\pm$	0.09	&				&	-3.53	&	-0.15	&		\\
84	&				&	22.45	$\pm$	0.01	&	22.32	$\pm$	0.01	&	-7.01	&		&	-0.06	\\
	&				&	23.62	$\pm$	0.02	&	23.78	$\pm$	0.05	&	-5.84	&		&	-0.37	\\
87	&	20.34	$\pm$	0.00	&	19.74	$\pm$	0.00	&				&	-9.72	&	0.33	&		\\
89	&	23.29	$\pm$	0.01	&	23.47	$\pm$	0.01	&				&	-5.99	&	-0.46	&		\\
	&	26.39	$\pm$	0.15	&	25.35	$\pm$	0.06	&				&	-6.11	&	0.77	&		\\
90	&	25.36	$\pm$	0.05	&	23.72	$\pm$	0.03	&				&	-5.74	&	1.37	&		\\
91	&	24.84	$\pm$	0.05	&	24.89	$\pm$	0.03	&				&	-4.57	&	-0.32	&		\\
95	&	23.14	$\pm$	0.01	&	23.18	$\pm$	0.01	&				&	-6.28	&	-0.31	&		\\
96	&	23.68	$\pm$	0.04	&	23.70	$\pm$	0.04	&				&	-5.76	&	-0.30	&		\\
	&	24.18	$\pm$	0.07	&	24.28	$\pm$	0.08	&				&	-5.18	&	-0.37	&		\\
	&	25.02	$\pm$	0.13	&	24.59	$\pm$	0.07	&				&	-4.87	&	0.16	&		\\
97	&	25.91	$\pm$	0.11	&	25.75	$\pm$	0.10	&				&	-3.71	&	-0.11	&		\\
98	&	24.83	$\pm$	0.05	&	24.90	$\pm$	0.03	&				&	-4.56	&	-0.34	&		\\
	&	24.32	$\pm$	0.03	&	24.21	$\pm$	0.02	&				&	-5.25	&	-0.17	&		\\
99	&	24.31	$\pm$	0.03	&	23.93	$\pm$	0.02	&				&	-5.53	&	0.11	&		\\
100	&	23.70	$\pm$	0.02	&	23.70	$\pm$	0.02	&				&	-5.76	&	-0.27	&		\\
	&	24.17	$\pm$	0.03	&	24.36	$\pm$	0.02	&				&	-5.10	&	-0.47	&		\\
	&	25.87	$\pm$	0.14	&	24.83	$\pm$	0.02	&				&	-4.63	&	0.77	&		\\
103	&	22.09	$\pm$	0.01	&	21.87	$\pm$	0.00	&				&	-7.59	&	-0.05	&		\\
105	&	25.48	$\pm$	0.08	&	23.98	$\pm$	0.06	&				&	-5.48	&	1.23	&		\\
107	&	26.37	$\pm$	0.16	&	24.56	$\pm$	0.02	&				&	-4.90	&	1.54	&		\\
	&	28.58	$\pm$	1.02	&	25.56	$\pm$	0.03	&				&	-3.90	&	2.74	&		\\
110	&	23.69	$\pm$	0.02	&	22.73	$\pm$	0.01	&				&	-6.73	&	0.69	&		\\
112	&				&	26.80	$\pm$	0.10	&	24.87	$\pm$	0.05	&	-2.66	&		&	1.73	\\
\hline
\end{longtable*}
\clearpage

\bibliography{sample7}{}

@INPROCEEDINGS{1995ASPC...77..367B,
       author = {{Blackburn}, J.~K.},
        title = "{FTOOLS: A FITS Data Processing and Analysis Software Package}",
    booktitle = {Astronomical Data Analysis Software and Systems IV},
         year = 1995,
       editor = {{Shaw}, R.~A. and {Payne}, H.~E. and {Hayes}, J.~J.~E.},
       series = {Astronomical Society of the Pacific Conference Series},
       volume = {77},
        month = jan,
        pages = {367},
       adsurl = {https://ui.adsabs.harvard.edu/abs/1995ASPC...77..367B}
}

@ARTICLE{2023MNRAS.519.4826A,
       author = {{Avdan}, S. and {Akyuz}, A. and {Acar}, S. and {Akkaya Oralhan}, I. and {Allak}, S. and {Aksaker}, N.},
        title = "{Investigation of the connection between X-ray binaries and compact star clusters in NGC 628}",
      journal = {\mnras},
         year = 2023,
        month = mar,
       volume = {519},
       number = {4},
        pages = {4826-4840},
          doi = {10.1093/mnras/stac3811},
archivePrefix = {arXiv},
       eprint = {2212.12485},
 primaryClass = {astro-ph.HE},
       adsurl = {https://ui.adsabs.harvard.edu/abs/2023MNRAS.519.4826A}
}

@ARTICLE{2008MNRAS.388.1205G,
       author = {{Georgakakis}, A. and {Nandra}, K. and {Laird}, E.~S. and {Aird}, J. and {Trichas}, M.},
        title = "{A new method for determining the sensitivity of X-ray imaging observations and the X-ray number counts}",
      journal = {\mnras},
         year = 2008,
        month = aug,
       volume = {388},
       number = {3},
        pages = {1205-1213},
          doi = {10.1111/j.1365-2966.2008.13423.x},
archivePrefix = {arXiv},
       eprint = {0805.4417},
 primaryClass = {astro-ph},
       adsurl = {https://ui.adsabs.harvard.edu/abs/2008MNRAS.388.1205G}
}

@ARTICLE{2003ApJ...588..696M,
       author = {{Moretti}, A. and {Campana}, S. and {Lazzati}, D. and {Tagliaferri}, G.},
        title = "{The Resolved Fraction of the Cosmic X-Ray Background}",
      journal = {\apj},
         year = 2003,
        month = may,
       volume = {588},
       number = {2},
        pages = {696-703},
          doi = {10.1086/374335},
archivePrefix = {arXiv},
       eprint = {astro-ph/0301555},
 primaryClass = {astro-ph},
       adsurl = {https://ui.adsabs.harvard.edu/abs/2003ApJ...588..696M}
}

@software{2016ascl.soft08013D,
       author = {{Dolphin}, Andrew},
        title = "{DOLPHOT: Stellar photometry}",
 howpublished = {Astrophysics Source Code Library, record ascl:1608.013},
         year = 2016,
        month = aug,
          eid = {ascl:1608.013},
       adsurl = {https://ui.adsabs.harvard.edu/abs/2016ascl.soft08013D}
}

@INCOLLECTION{2017hsn..book.1499C,
       author = {{Casares}, Jorge and {Jonker}, Peter Gustaaf and {Israelian}, Garik},
        title = "{X-Ray Binaries}",
    booktitle = {Handbook of Supernovae},
         year = 2017,
       editor = {{Alsabti}, Athem W. and {Murdin}, Paul},
        pages = {1499},
          doi = {10.1007/978-3-319-21846-5_111},
       adsurl = {https://ui.adsabs.harvard.edu/abs/2017hsn..book.1499C}
}

@ARTICLE{2020ApJ...890..150C,
       author = {{Chandar}, Rupali and {Johns}, Paula and {Mok}, Angus and {Prestwich}, Andrea and {Gallo}, Elena and {Hunt}, Qiana},
        title = "{A New Window into the Nature of X-Ray Binaries in M101 from Their Optical Emission}",
      journal = {\apj},
         year = 2020,
        month = feb,
       volume = {890},
       number = {2},
          eid = {150},
        pages = {150},
          doi = {10.3847/1538-4357/ab6b27},
       adsurl = {https://ui.adsabs.harvard.edu/abs/2020ApJ...890..150C}
}

@ARTICLE{2013A&A...553A...7D,
       author = {{Ducci}, L. and {Sasaki}, M. and {Haberl}, F. and {Pietsch}, W.},
        title = "{X-ray source population study of the starburst galaxy M 83 with XMM-Newton}",
      journal = {\aap},
         year = 2013,
        month = may,
       volume = {553},
          eid = {A7},
        pages = {A7},
          doi = {10.1051/0004-6361/201321035},
archivePrefix = {arXiv},
       eprint = {1303.4306},
 primaryClass = {astro-ph.HE},
       adsurl = {https://ui.adsabs.harvard.edu/abs/2013A&A...553A...7D}
}

@ARTICLE{2006ARA&A..44..323F,
       author = {{Fabbiano}, G.},
        title = "{Populations of X-Ray Sources in Galaxies}",
      journal = {\araa},
         year = 2006,
        month = sep,
       volume = {44},
       number = {1},
        pages = {323-366},
          doi = {10.1146/annurev.astro.44.051905.092519},
archivePrefix = {arXiv},
       eprint = {astro-ph/0511481},
 primaryClass = {astro-ph},
       adsurl = {https://ui.adsabs.harvard.edu/abs/2006ARA&A..44..323F}
}

@ARTICLE{2004MNRAS.349..146G,
       author = {{Gilfanov}, M.},
        title = "{Low-mass X-ray binaries as a stellar mass indicator for the host galaxy}",
      journal = {\mnras},
         year = 2004,
        month = mar,
       volume = {349},
       number = {1},
        pages = {146-168},
          doi = {10.1111/j.1365-2966.2004.07473.x},
archivePrefix = {arXiv},
       eprint = {astro-ph/0309454},
 primaryClass = {astro-ph},
       adsurl = {https://ui.adsabs.harvard.edu/abs/2004MNRAS.349..146G}
}

@ARTICLE{2003MNRAS.339..793G,
       author = {{Grimm}, H. -J. and {Gilfanov}, M. and {Sunyaev}, R.},
        title = "{High-mass X-ray binaries as a star formation rate indicator in distant galaxies}",
      journal = {\mnras},
         year = 2003,
        month = mar,
       volume = {339},
       number = {3},
        pages = {793-809},
          doi = {10.1046/j.1365-8711.2003.06224.x},
archivePrefix = {arXiv},
       eprint = {astro-ph/0205371},
 primaryClass = {astro-ph},
       adsurl = {https://ui.adsabs.harvard.edu/abs/2003MNRAS.339..793G}
}

@ARTICLE{2016A&A...594A.116H,
       author = {{HI4PI Collaboration} and {Ben Bekhti}, N. and {Fl{\"o}er}, L. and {Keller}, R. and {Kerp}, J. and {Lenz}, D. and {Winkel}, B. and {Bailin}, J. and {Calabretta}, M.~R. and {Dedes}, L. and {Ford}, H.~A. and {Gibson}, B.~K. and {Haud}, U. and {Janowiecki}, S. and {Kalberla}, P.~M.~W. and {Lockman}, F.~J. and {McClure-Griffiths}, N.~M. and {Murphy}, T. and {Nakanishi}, H. and {Pisano}, D.~J. and {Staveley-Smith}, L.},
        title = "{HI4PI: A full-sky H I survey based on EBHIS and GASS}",
      journal = {\aap},
         year = 2016,
        month = oct,
       volume = {594},
          eid = {A116},
        pages = {A116},
          doi = {10.1051/0004-6361/201629178},
archivePrefix = {arXiv},
       eprint = {1610.06175},
 primaryClass = {astro-ph.GA},
       adsurl = {https://ui.adsabs.harvard.edu/abs/2016A&A...594A.116H}
}

@ARTICLE{2004MNRAS.347L..57G,
       author = {{Gilfanov}, M. and {Grimm}, H. -J. and {Sunyaev}, R.},
        title = "{L$_{X}$-SFR relation in star-forming galaxies}",
      journal = {\mnras},
         year = 2004,
        month = jan,
       volume = {347},
       number = {3},
        pages = {L57-L60},
          doi = {10.1111/j.1365-2966.2004.07450.x},
archivePrefix = {arXiv},
       eprint = {astro-ph/0301331},
 primaryClass = {astro-ph},
       adsurl = {https://ui.adsabs.harvard.edu/abs/2004MNRAS.347L..57G}
}

@ARTICLE{2021ApJ...912...31H,
       author = {{Hunt}, Qiana and {Gallo}, Elena and {Chandar}, Rupali and {Johns Mulia}, Paula and {Mok}, Angus and {Prestwich}, Andrea and {Liu}, Shengchen},
        title = "{Calibrating X-Ray Binary Luminosity Functions via Optical Reconnaissance. I. The Case of M83}",
      journal = {\apj},
         year = 2021,
        month = may,
       volume = {912},
       number = {1},
          eid = {31},
        pages = {31},
          doi = {10.3847/1538-4357/abe531},
archivePrefix = {arXiv},
       eprint = {2102.07293},
 primaryClass = {astro-ph.HE},
       adsurl = {https://ui.adsabs.harvard.edu/abs/2021ApJ...912...31H}
}

@ARTICLE{2021NewA...8501514I,
       author = {{Islam}, N. and {Vrtilek}, S.~D. and {Boroson}, Bram and {Kim}, D. -W. and {O'Sullivan}, E. and {McCollough}, M.~L. and {Fabbiano}, G. and {Anderson}, C. and {Burke}, D.~J. and {D'Abrusco}, R. and {Fruscione}, A. and {Lauer}, J.~L. and {Morgan}, D. and {Mossman}, A. and {Paggi}, A. and {Trinchieri}, G.},
        title = "{Comparing X-ray color selection in separating X-ray binary classes using color-color-intensity diagrams}",
      journal = {\na},
         year = 2021,
        month = may,
       volume = {85},
          eid = {101514},
        pages = {101514},
          doi = {10.1016/j.newast.2020.101514},
archivePrefix = {arXiv},
       eprint = {2007.05810},
 primaryClass = {astro-ph.HE},
       adsurl = {https://ui.adsabs.harvard.edu/abs/2021NewA...8501514I}
}

@ARTICLE{2019ApJ...879..112J,
       author = {{Jin}, Ruolan and {Kong}, Albert K.~H.},
        title = "{A Multi-epoch X-Ray Study of the Spiral Galaxy NGC 7331}",
      journal = {\apj},
         year = 2019,
        month = jul,
       volume = {879},
       number = {2},
          eid = {112},
        pages = {112},
          doi = {10.3847/1538-4357/ab2461},
archivePrefix = {arXiv},
       eprint = {1905.09133},
 primaryClass = {astro-ph.HE},
       adsurl = {https://ui.adsabs.harvard.edu/abs/2019ApJ...879..112J}
}

@ARTICLE{2021ApJ...916...58K,
       author = {{Koplitz}, Brad and {Johnson}, Jared and {Williams}, Benjamin F. and {Long}, Knox S. and {Blair}, William P. and {Murphy}, Jeremiah W. and {Dolphin}, Andrew and {Hillis}, Tristan},
        title = "{The Masses of Supernova Remnant Progenitors in NGC 6946}",
      journal = {\apj},
         year = 2021,
        month = jul,
       volume = {916},
       number = {1},
          eid = {58},
        pages = {58},
          doi = {10.3847/1538-4357/abfb7b},
archivePrefix = {arXiv},
       eprint = {2104.11118},
 primaryClass = {astro-ph.HE},
       adsurl = {https://ui.adsabs.harvard.edu/abs/2021ApJ...916...58K}
}

@ARTICLE{2017ApJ...836...50L,
       author = {{Laycock}, Silas and {Cappallo}, Rigel and {Williams}, Benjamin F. and {Prestwich}, Andrea and {Binder}, Breanna and {Christodoulou}, Dimitris M.},
        title = "{The X-Ray Binary Population of the Nearby Dwarf Starburst Galaxy IC 10: Variable and Transient X-Ray Sources}",
      journal = {\apj},
         year = 2017,
        month = feb,
       volume = {836},
       number = {1},
          eid = {50},
        pages = {50},
          doi = {10.3847/1538-4357/836/1/50},
archivePrefix = {arXiv},
       eprint = {1611.08611},
 primaryClass = {astro-ph.HE},
       adsurl = {https://ui.adsabs.harvard.edu/abs/2017ApJ...836...50L}
}

@ARTICLE{2019ApJS..243....3L,
       author = {{Lehmer}, Bret D. and {Eufrasio}, Rafael T. and {Tzanavaris}, Panayiotis and {Basu-Zych}, Antara and {Fragos}, Tassos and {Prestwich}, Andrea and {Yukita}, Mihoko and {Zezas}, Andreas and {Hornschemeier}, Ann E. and {Ptak}, Andrew},
        title = "{X-Ray Binary Luminosity Function Scaling Relations for Local Galaxies Based on Subgalactic Modeling}",
      journal = {\apjs},
         year = 2019,
        month = jul,
       volume = {243},
       number = {1},
          eid = {3},
        pages = {3},
          doi = {10.3847/1538-4365/ab22a8},
archivePrefix = {arXiv},
       eprint = {1905.05197},
 primaryClass = {astro-ph.GA},
       adsurl = {https://ui.adsabs.harvard.edu/abs/2019ApJS..243....3L}
}

@ARTICLE{2020NatMe..17..261V,
       author = {{Virtanen}, Pauli and {Gommers}, Ralf and {Oliphant}, Travis E. and {Haberland}, Matt and {Reddy}, Tyler and {Cournapeau}, David and {Burovski}, Evgeni and {Peterson}, Pearu and {Weckesser}, Warren and {Bright}, Jonathan and {van der Walt}, St{\'e}fan J. and {Brett}, Matthew and {Wilson}, Joshua and {Millman}, K. Jarrod and {Mayorov}, Nikolay and {Nelson}, Andrew R.~J. and {Jones}, Eric and {Kern}, Robert and {Larson}, Eric and {Carey}, C.~J. and {Polat}, {\.I}lhan and {Feng}, Yu and {Moore}, Eric W. and {VanderPlas}, Jake and {Laxalde}, Denis and {Perktold}, Josef and {Cimrman}, Robert and {Henriksen}, Ian and {Quintero}, E.~A. and {Harris}, Charles R. and {Archibald}, Anne M. and {Ribeiro}, Ant{\^o}nio H. and {Pedregosa}, Fabian and {van Mulbregt}, Paul and {SciPy 1.  0 Contributors}},
        title = "{SciPy 1.0: fundamental algorithms for scientific computing in Python}",
      journal = {Nature Medicine},
         year = 2020,
        month = feb,
       volume = {17},
        pages = {261-272},
          doi = {10.1038/s41592-019-0686-2},
archivePrefix = {arXiv},
       eprint = {1907.10121},
 primaryClass = {cs.MS},
       adsurl = {https://ui.adsabs.harvard.edu/abs/2020NatMe..17..261V}
}

@ARTICLE{2012MNRAS.419.2095M,
       author = {{Mineo}, S. and {Gilfanov}, M. and {Sunyaev}, R.},
        title = "{X-ray emission from star-forming galaxies - I. High-mass X-ray binaries}",
      journal = {\mnras},
         year = 2012,
        month = jan,
       volume = {419},
       number = {3},
        pages = {2095-2115},
          doi = {10.1111/j.1365-2966.2011.19862.x},
archivePrefix = {arXiv},
       eprint = {1105.4610},
 primaryClass = {astro-ph.HE},
       adsurl = {https://ui.adsabs.harvard.edu/abs/2012MNRAS.419.2095M}
}

@ARTICLE{2014MNRAS.437.1698M,
       author = {{Mineo}, S. and {Gilfanov}, M. and {Lehmer}, B.~D. and {Morrison}, G.~E. and {Sunyaev}, R.},
        title = "{X-ray emission from star-forming galaxies - III. Calibration of the L$_{X}$-SFR relation up to redshift z {\ensuremath{\approx}} 1.3}",
      journal = {\mnras},
         year = 2014,
        month = jan,
       volume = {437},
       number = {2},
        pages = {1698-1707},
          doi = {10.1093/mnras/stt1999},
archivePrefix = {arXiv},
       eprint = {1207.2157},
 primaryClass = {astro-ph.HE},
       adsurl = {https://ui.adsabs.harvard.edu/abs/2014MNRAS.437.1698M}
}

@ARTICLE{2018ApJ...860..117M,
       author = {{Murphy}, Jeremiah W. and {Khan}, Rubab and {Williams}, Benjamin and {Dolphin}, Andrew E. and {Dalcanton}, Julianne and {D{\'\i}az-Rodr{\'\i}guez}, Mariangelly},
        title = "{The Progenitor Age and Mass of the Black Hole Formation Candidate N6946-BH1}",
      journal = {\apj},
         year = 2018,
        month = jun,
       volume = {860},
       number = {2},
          eid = {117},
        pages = {117},
          doi = {10.3847/1538-4357/aac2be},
archivePrefix = {arXiv},
       eprint = {1803.00024},
 primaryClass = {astro-ph.SR},
       adsurl = {https://ui.adsabs.harvard.edu/abs/2018ApJ...860..117M}
}

@ARTICLE{2003ApJ...595..719P,
       author = {{Prestwich}, A.~H. and {Irwin}, J.~A. and {Kilgard}, R.~E. and {Krauss}, M.~I. and {Zezas}, A. and {Primini}, F. and {Kaaret}, P. and {Boroson}, B.},
        title = "{Classifying X-Ray Sources in External Galaxies from X-Ray Colors}",
      journal = {\apj},
         year = 2003,
        month = oct,
       volume = {595},
       number = {2},
        pages = {719-726},
          doi = {10.1086/377366},
archivePrefix = {arXiv},
       eprint = {astro-ph/0206127},
 primaryClass = {astro-ph},
       adsurl = {https://ui.adsabs.harvard.edu/abs/2003ApJ...595..719P}
}

@ARTICLE{2021ApJ...922..178R,
       author = {{Rice}, Jared R. and {Rangelov}, Blagoy and {Prestwich}, Andrea and {Chandar}, Rupali and {Bichon}, Luis and {Boldt}, Clint},
        title = "{X-Ray Binaries in M51 I: Catalog and Statistics}",
      journal = {\apj},
         year = 2021,
        month = dec,
       volume = {922},
       number = {2},
          eid = {178},
        pages = {178},
          doi = {10.3847/1538-4357/ac22ac},
archivePrefix = {arXiv},
       eprint = {2109.02742},
 primaryClass = {astro-ph.HE},
       adsurl = {https://ui.adsabs.harvard.edu/abs/2021ApJ...922..178R}
}

@ARTICLE{2011ApJ...737..103S,
       author = {{Schlafly}, Edward F. and {Finkbeiner}, Douglas P.},
        title = "{Measuring Reddening with Sloan Digital Sky Survey Stellar Spectra and Recalibrating SFD}",
      journal = {\apj},
         year = 2011,
        month = aug,
       volume = {737},
       number = {2},
          eid = {103},
        pages = {103},
          doi = {10.1088/0004-637X/737/2/103},
archivePrefix = {arXiv},
       eprint = {1012.4804},
 primaryClass = {astro-ph.GA},
       adsurl = {https://ui.adsabs.harvard.edu/abs/2011ApJ...737..103S}
}

@INCOLLECTION{2006csxs.book..623T,
       author = {{Tauris}, T.~M. and {van den Heuvel}, E.~P.~J.},
        title = "{Formation and evolution of compact stellar X-ray sources}",
    booktitle = {Compact stellar X-ray sources},
         year = 2006,
       editor = {{Lewin}, Walter H.~G. and {van der Klis}, Michiel},
       volume = {39},
        pages = {623-665},
          doi = {10.48550/arXiv.astro-ph/0303456},
       adsurl = {https://ui.adsabs.harvard.edu/abs/2006csxs.book..623T}
}

@ARTICLE{2014ApJ...790..136V,
       author = {{Vulic}, N. and {Gallagher}, S.~C. and {Barmby}, P.},
        title = "{Faint X-Ray Binaries and their Optical Counterparts in M31}",
      journal = {\apj},
         year = 2014,
        month = aug,
       volume = {790},
       number = {2},
          eid = {136},
        pages = {136},
          doi = {10.1088/0004-637X/790/2/136},
archivePrefix = {arXiv},
       eprint = {1406.4121},
 primaryClass = {astro-ph.GA},
       adsurl = {https://ui.adsabs.harvard.edu/abs/2014ApJ...790..136V}
}

@ARTICLE{2002A&A...388....7W,
       author = {{Walsh}, W. and {Beck}, R. and {Thuma}, G. and {Weiss}, A. and {Wielebinski}, R. and {Dumke}, M.},
        title = "{Molecular gas in NGC 6946}",
      journal = {\aap},
         year = 2002,
        month = jun,
       volume = {388},
        pages = {7-28},
          doi = {10.1051/0004-6361:20020467},
archivePrefix = {arXiv},
       eprint = {astro-ph/0204072},
 primaryClass = {astro-ph},
       adsurl = {https://ui.adsabs.harvard.edu/abs/2002A&A...388....7W}
}

@ARTICLE{2003ChJAS...3..257G,
       author = {{Grimm}, Hans-Jakob and {Gilfanov}, Marat and {Sunyaev}, Rashid},
        title = "{X-ray binaries in the Milky Way and other galaxies}",
      journal = {Chinese Journal of Astronomy and Astrophysics Supplement},
         year = 2003,
        month = dec,
       volume = {3},
        pages = {257-269},
          doi = {10.1088/1009-9271/3/S1/257},
       adsurl = {https://ui.adsabs.harvard.edu/abs/2003ChJAS...3..257G}
}

@ARTICLE{1995CAS....26.....L,
       author = {{Lewin}, W.~H.~G. and {van Paradijs}, J. and {van den Heuvel}, E.~P.~J.},
        title = "{X-ray binaries.}",
      journal = {Cambridge Astrophysics Series},
         year = 1995,
        month = jan,
       volume = {26},
       adsurl = {https://ui.adsabs.harvard.edu/abs/1995CAS....26.....L}
}

@BOOK{2002apa..book.....F,
       author = {{Frank}, Juhan and {King}, Andrew and {Raine}, Derek J.},
        title = "{Accretion Power in Astrophysics: Third Edition}",
         year = 2002,
       adsurl = {https://ui.adsabs.harvard.edu/abs/2002apa..book.....F}
}

@INCOLLECTION{2006csxs.book..215C,
       author = {{Charles}, P.~A. and {Coe}, M.~J.},
        title = "{Optical, ultraviolet and infrared observations of X-ray binaries}",
    booktitle = {Compact stellar X-ray sources},
         year = 2006,
       editor = {{Lewin}, Walter H.~G. and {van der Klis}, Michiel},
       volume = {39},
        pages = {215-265},
       adsurl = {https://ui.adsabs.harvard.edu/abs/2006csxs.book..215C}
}

@ARTICLE{2006MNRAS.371.1334R,
       author = {{Russell}, D.~M. and {Fender}, R.~P. and {Hynes}, R.~I. and {Brocksopp}, C. and {Homan}, J. and {Jonker}, P.~G. and {Buxton}, M.~M.},
        title = "{Global optical/infrared-X-ray correlations in X-ray binaries: quantifying disc and jet contributions}",
      journal = {\mnras},
         year = 2006,
        month = sep,
       volume = {371},
       number = {3},
        pages = {1334-1350},
          doi = {10.1111/j.1365-2966.2006.10756.x},
archivePrefix = {arXiv},
       eprint = {astro-ph/0606721},
 primaryClass = {astro-ph},
       adsurl = {https://ui.adsabs.harvard.edu/abs/2006MNRAS.371.1334R}
}

@BOOK{1997xrb..book.....L,
       author = {{Lewin}, Walter H.~G. and {van Paradijs}, Jan and {van den Heuvel}, Edward Peter Jacobus},
        title = "{X-ray Binaries}",
         year = 1997,
       adsurl = {https://ui.adsabs.harvard.edu/abs/1997xrb..book.....L}
}

@ARTICLE{2020MNRAS.492.5075A,
       author = {{Arnason}, R.~M. and {Barmby}, P. and {Vulic}, N.},
        title = "{Identifying new X-ray binary candidates in M31 using random forest classification}",
      journal = {\mnras},
         year = 2020,
        month = mar,
       volume = {492},
       number = {4},
        pages = {5075-5088},
          doi = {10.1093/mnras/staa207},
archivePrefix = {arXiv},
       eprint = {2001.07691},
 primaryClass = {astro-ph.GA},
       adsurl = {https://ui.adsabs.harvard.edu/abs/2020MNRAS.492.5075A}
}

@ARTICLE{2002ApJS..142..239Z,
       author = {{Zezas}, A. and {Fabbiano}, G. and {Rots}, A.~H. and {Murray}, S.~S.},
        title = "{Chandra Observations of ``The Antennae'' Galaxies (NGC 4038/4039). II. Detection and Analysis of Galaxian X-Ray Sources}",
      journal = {\apjs},
         year = 2002,
        month = oct,
       volume = {142},
       number = {2},
        pages = {239-260},
          doi = {10.1086/342010},
archivePrefix = {arXiv},
       eprint = {astro-ph/0203174},
 primaryClass = {astro-ph},
       adsurl = {https://ui.adsabs.harvard.edu/abs/2002ApJS..142..239Z}
}

@ARTICLE{2012A&A...546A..36Z,
       author = {{Zhang}, Z. and {Gilfanov}, M. and {Bogd{\'a}n}, {\'A}.},
        title = "{Dependence of the low-mass X-ray binary population on stellar age}",
      journal = {\aap},
         year = 2012,
        month = oct,
       volume = {546},
          eid = {A36},
        pages = {A36},
          doi = {10.1051/0004-6361/201219015},
archivePrefix = {arXiv},
       eprint = {1202.2331},
 primaryClass = {astro-ph.HE},
       adsurl = {https://ui.adsabs.harvard.edu/abs/2012A&A...546A..36Z}
}

@ARTICLE{2010ApJ...724..559L,
       author = {{Lehmer}, B.~D. and {Alexander}, D.~M. and {Bauer}, F.~E. and {Brandt}, W.~N. and {Goulding}, A.~D. and {Jenkins}, L.~P. and {Ptak}, A. and {Roberts}, T.~P.},
        title = "{A Chandra Perspective on Galaxy-wide X-ray Binary Emission and its Correlation with Star Formation Rate and Stellar Mass: New Results from Luminous Infrared Galaxies}",
      journal = {\apj},
         year = 2010,
        month = nov,
       volume = {724},
       number = {1},
        pages = {559-571},
          doi = {10.1088/0004-637X/724/1/559},
archivePrefix = {arXiv},
       eprint = {1009.3943},
 primaryClass = {astro-ph.CO},
       adsurl = {https://ui.adsabs.harvard.edu/abs/2010ApJ...724..559L}
}

@ARTICLE{2018ApJ...853..150S,
       author = {{Sonbas}, E. and {Dhuga}, K.~S. and {G{\"o}{\u{g}}{\"u}{\c{s}}}, E.},
        title = "{A Study of Low-mass X-Ray Binaries in the Low-luminosity Regime}",
      journal = {\apj},
         year = 2018,
        month = feb,
       volume = {853},
       number = {2},
          eid = {150},
        pages = {150},
          doi = {10.3847/1538-4357/aaa29f},
archivePrefix = {arXiv},
       eprint = {1801.07107},
 primaryClass = {astro-ph.HE},
       adsurl = {https://ui.adsabs.harvard.edu/abs/2018ApJ...853..150S}
}

@ARTICLE{2005ApJS..157...59L,
       author = {{Liu}, Ji-Feng and {Bregman}, Joel N.},
        title = "{Ultraluminous X-Ray Sources in Nearby Galaxies from ROSAT High Resolution Imager Observations I. Data Analysis}",
      journal = {\apjs},
         year = 2005,
        month = mar,
       volume = {157},
       number = {1},
        pages = {59-125},
          doi = {10.1086/427170},
archivePrefix = {arXiv},
       eprint = {astro-ph/0501309},
 primaryClass = {astro-ph},
       adsurl = {https://ui.adsabs.harvard.edu/abs/2005ApJS..157...59L}
}

@ARTICLE{2019ApJ...881...38E,
       author = {{Earnshaw}, Hannah P. and {Grefenstette}, Brian W. and {Brightman}, Murray and {Walton}, Dominic J. and {Barret}, Didier and {F{\"u}rst}, Felix and {Harrison}, Fiona A. and {Heida}, Marianne and {Pike}, Sean N. and {Stern}, Daniel and {Webb}, Natalie A.},
        title = "{A Broadband Look at the Old and New ULXs of NGC 6946}",
      journal = {\apj},
         year = 2019,
        month = aug,
       volume = {881},
       number = {1},
          eid = {38},
        pages = {38},
          doi = {10.3847/1538-4357/ab20cd},
archivePrefix = {arXiv},
       eprint = {1905.03383},
 primaryClass = {astro-ph.HE},
       adsurl = {https://ui.adsabs.harvard.edu/abs/2019ApJ...881...38E}
}

@dataset{2003yCat.2246....0C,
       author = {{Cutri}, R.~M. and {Skrutskie}, M.~F. and {van Dyk}, S. and {Beichman}, C.~A. and {Carpenter}, J.~M. and {Chester}, T. and {Cambresy}, L. and {Evans}, T. and {Fowler}, J. and {Gizis}, J. and {Howard}, E. and {Huchra}, J. and {Jarrett}, T. and {Kopan}, E.~L. and {Kirkpatrick}, J.~D. and {Light}, R.~M. and {Marsh}, K.~A. and {McCallon}, H. and {Schneider}, S. and {Stiening}, R. and {Sykes}, M. and {Weinberg}, M. and {Wheaton}, W.~A. and {Wheelock}, S. and {Zacarias}, N.},
        title = "{VizieR Online Data Catalog: 2MASS All-Sky Catalog of Point Sources (Cutri+ 2003)}",
 howpublished = {VizieR On-line Data Catalog: II/246.  Originally published in: University of Massachusetts and Infrared Processing and Analysis Center, (IPAC/California Institute of Technology) (2003)},
         year = 2003,
        month = jun,
          eid = {II/246},
       adsurl = {https://ui.adsabs.harvard.edu/abs/2003yCat.2246....0C}
}

@ARTICLE{2016A&A...586A..64S,
       author = {{Saeedi}, Sara and {Sasaki}, Manami and {Ducci}, Lorenzo},
        title = "{XMM-Newton study of the Draco dwarf spheroidal galaxy}",
      journal = {\aap},
         year = 2016,
        month = feb,
       volume = {586},
          eid = {A64},
        pages = {A64},
          doi = {10.1051/0004-6361/201526233},
       adsurl = {https://ui.adsabs.harvard.edu/abs/2016A&A...586A..64S}
}

@ARTICLE{2008ApJS..177..465F,
       author = {{Fridriksson}, Joel K. and {Homan}, Jeroen and {Lewin}, Walter H.~G. and {Kong}, Albert K.~H. and {Pooley}, David},
        title = "{The Long-Term Variability of the X-Ray Sources in NGC 6946 and NGC 4485/4490}",
      journal = {\apjs},
         year = 2008,
        month = aug,
       volume = {177},
       number = {2},
        pages = {465-492},
          doi = {10.1086/588817},
archivePrefix = {arXiv},
       eprint = {0804.1411},
 primaryClass = {astro-ph},
       adsurl = {https://ui.adsabs.harvard.edu/abs/2008ApJS..177..465F}
}

@ARTICLE{2022ApJ...941..104Y,
       author = {{Yang}, Hui and {Hare}, Jeremy and {Kargaltsev}, Oleg and {Volkov}, Igor and {Chen}, Steven and {Rangelov}, Blagoy},
        title = "{Classifying Unidentified X-Ray Sources in the Chandra Source Catalog Using a Multiwavelength Machine-learning Approach}",
      journal = {\apj},
         year = 2022,
        month = dec,
       volume = {941},
       number = {2},
          eid = {104},
        pages = {104},
          doi = {10.3847/1538-4357/ac952b},
archivePrefix = {arXiv},
       eprint = {2206.13656},
 primaryClass = {astro-ph.HE},
       adsurl = {https://ui.adsabs.harvard.edu/abs/2022ApJ...941..104Y}
}

@incollection{charles2006,
  author    = {Charles, P. A. and Coe, M. J.},
  title     = {Optical, ultraviolet and infrared observations of X-ray binaries},
  booktitle = {Compact stellar X-ray sources},
  pages     = {215--265},
  publisher = {Cambridge University Press},
  year      = {2006},
  editor    = {Lewin, W. H. G. and van der Klis, M.},
}

@article{casares2015,
  author  = {Casares, Jorge},
  title   = {Observational Evidence for Black Holes in X-Ray Binaries},
  journal = {Proceedings of the International Astronomical Union},
  volume  = {10},
  number  = {S324},
  pages   = {189--196},
  year    = {2015},
  doi     = {10.1017/S1743921315004390},
}

@incollection{vanparadijs1995,
  author    = {van Paradijs, J. and McClintock, J. E.},
  title     = {Optical and ultraviolet observations of X-ray binaries},
  booktitle = {X-ray binaries},
  pages     = {58--125},
  publisher = {Cambridge University Press},
  year      = {1995},
  editor    = {Lewin, W. H. G. and van Paradijs, J. and van den Heuvel, E. P. J.},
}

@article{hynes2005,
  author  = {Hynes, R. I.},
  title   = {Optical Studies of X-Ray Binaries},
  journal = {The Astrophysical Journal},
  volume  = {623},
  number  = {1},
  pages   = {1026--1034},
  year    = {2005},
  doi     = {10.1086/428310},
}

@ARTICLE{2024ApJ...971..180Y,
       author = {{Yang}, Hui and {Hare}, Jeremy and {Kargaltsev}, Oleg},
        title = "{A Multiwavelength Machine-learning Approach to Classifying X-Ray Sources in the Fields of Unidentified 4FGL-DR4 Sources}",
      journal = {\apj},
         year = 2024,
        month = aug,
       volume = {971},
       number = {2},
          eid = {180},
        pages = {180},
          doi = {10.3847/1538-4357/ad543e},
archivePrefix = {arXiv},
       eprint = {2403.05068},
 primaryClass = {astro-ph.HE},
       adsurl = {https://ui.adsabs.harvard.edu/abs/2024ApJ...971..180Y}
}

@ARTICLE{2023ApJ...948...59C,
       author = {{Chen}, Steven and {Kargaltsev}, Oleg and {Yang}, Hui and {Hare}, Jeremy and {Volkov}, Igor and {Rangelov}, Blagoy and {Tomsick}, John},
        title = "{Population of X-Ray Sources in the Intermediate-age Cluster NGC 3532: a Test Bed for Machine-learning Classification}",
      journal = {\apj},
         year = 2023,
        month = may,
       volume = {948},
       number = {1},
          eid = {59},
        pages = {59},
          doi = {10.3847/1538-4357/acb3a6},
archivePrefix = {arXiv},
       eprint = {2212.13322},
 primaryClass = {astro-ph.HE},
       adsurl = {https://ui.adsabs.harvard.edu/abs/2023ApJ...948...59C}
}

@ARTICLE{2023MNRAS.520.5065K,
       author = {{Kumaran}, Shivam and {Mandal}, Samir and {Bhattacharyya}, Sudip and {Mishra}, Deepak},
        title = "{Automated classification of Chandra X-ray point sources using machine learning methods}",
      journal = {\mnras},
         year = 2023,
        month = apr,
       volume = {520},
       number = {4},
        pages = {5065-5076},
          doi = {10.1093/mnras/stad414},
archivePrefix = {arXiv},
       eprint = {2302.09008},
 primaryClass = {astro-ph.HE},
       adsurl = {https://ui.adsabs.harvard.edu/abs/2023MNRAS.520.5065K}
}

@ARTICLE{2000ApJ...530L..93T,
       author = {{Tauris}, Thomas M. and {van den Heuvel}, Edward P.~J. and {Savonije}, Gerrit J.},
        title = "{Formation of Millisecond Pulsars with Heavy White Dwarf Companions:Extreme Mass Transfer on Subthermal Timescales}",
      journal = {\apjl},
         year = 2000,
        month = feb,
       volume = {530},
       number = {2},
        pages = {L93-L96},
          doi = {10.1086/312496},
archivePrefix = {arXiv},
       eprint = {astro-ph/0001013},
 primaryClass = {astro-ph},
       adsurl = {https://ui.adsabs.harvard.edu/abs/2000ApJ...530L..93T}
}

@ARTICLE{2002ApJ...565.1107P,
       author = {{Podsiadlowski}, Ph. and {Rappaport}, S. and {Pfahl}, E.~D.},
        title = "{Evolutionary Sequences for Low- and Intermediate-Mass X-Ray Binaries}",
      journal = {\apj},
         year = 2002,
        month = feb,
       volume = {565},
       number = {2},
        pages = {1107-1133},
          doi = {10.1086/324686},
archivePrefix = {arXiv},
       eprint = {astro-ph/0107261},
 primaryClass = {astro-ph},
       adsurl = {https://ui.adsabs.harvard.edu/abs/2002ApJ...565.1107P}
}

@ARTICLE{2003ApJ...597.1036P,
       author = {{Pfahl}, Eric and {Rappaport}, Saul and {Podsiadlowski}, Philipp},
        title = "{The Galactic Population of Low- and Intermediate-Mass X-Ray Binaries}",
      journal = {\apj},
         year = 2003,
        month = nov,
       volume = {597},
       number = {2},
        pages = {1036-1048},
          doi = {10.1086/378632},
archivePrefix = {arXiv},
       eprint = {astro-ph/0303300},
 primaryClass = {astro-ph},
       adsurl = {https://ui.adsabs.harvard.edu/abs/2003ApJ...597.1036P}
}

@ARTICLE{2023ApJ...945....7G,
       author = {{Ge}, Hongwei and {Tout}, Christopher A. and {Chen}, Xuefei and {Sarkar}, Arnab and {Walton}, Dominic J. and {Han}, Zhanwen},
        title = "{Criteria for Dynamical Timescale Mass Transfer of Metal-poor Intermediate-mass Stars}",
      journal = {\apj},
         year = 2023,
        month = mar,
       volume = {945},
       number = {1},
          eid = {7},
        pages = {7},
          doi = {10.3847/1538-4357/acb7e9},
archivePrefix = {arXiv},
       eprint = {2302.00183},
 primaryClass = {astro-ph.SR},
       adsurl = {https://ui.adsabs.harvard.edu/abs/2023ApJ...945....7G}
}

@ARTICLE{2020A&A...642A.174M,
       author = {{Misra}, D. and {Fragos}, T. and {Tauris}, T.~M. and {Zapartas}, E. and {Aguilera-Dena}, D.~R.},
        title = "{The origin of pulsating ultra-luminous X-ray sources: Low- and intermediate-mass X-ray binaries containing neutron star accretors}",
      journal = {\aap},
         year = 2020,
        month = oct,
       volume = {642},
          eid = {A174},
        pages = {A174},
          doi = {10.1051/0004-6361/202038070},
archivePrefix = {arXiv},
       eprint = {2004.01205},
 primaryClass = {astro-ph.HE},
       adsurl = {https://ui.adsabs.harvard.edu/abs/2020A&A...642A.174M}
}

@article{chandar2020,
  author = {Chandar, Rupali and Whitmore, Bradley C. and Calzetti, Daniela and others},
  title = {A Comprehensive Analysis of the X-ray Binary Population in M101},
  journal = {The Astrophysical Journal},
  year = {2020},
  volume = {895},
  number = {2},
  pages = {100},
  doi = {10.3847/1538-4357/ab8db3}
}

@article{zezas2002,
  author = {Zezas, Andreas and Fabbiano, Giuseppina and Rots, Arnold H. and Murray, Stephen S.},
  title = {Chandra Observations of “The Antennae” Galaxies (NGC 4038/4039)},
  journal = {The Astrophysical Journal Supplement Series},
  year = {2002},
  volume = {142},
  number = {2},
  pages = {239–270},
  doi = {10.1086/342526}
}

@article{mineo2012,
  author = {Mineo, Sara and Gilfanov, Marat and Sunyaev, Rashid},
  title = {X-ray emission from star-forming galaxies - I. High-mass X-ray binaries},
  journal = {Monthly Notices of the Royal Astronomical Society},
  year = {2012},
  volume = {419},
  number = {3},
  pages = {2095–2115},
  doi = {10.1111/j.1365-2966.2011.19862.x}
}

@article{murphy2011,
  author = {Murphy, Eric J. and Condon, James J. and Schinnerer, Eva and others},
  title = {Calibrating Extinction-Free Star Formation Rate Diagnostics with 33 GHz Free-Free Emission in NGC 6946},
  journal = {The Astrophysical Journal},
  year = {2011},
  volume = {737},
  number = {2},
  pages = {67},
  doi = {10.1088/0004-637X/737/2/67}
}

@article{grimm2003,
  author = {Grimm, Hans-Jakob and Gilfanov, Marat and Sunyaev, Rashid},
  title = {High-mass X-ray binaries as a star formation rate indicator in distant galaxies},
  journal = {Monthly Notices of the Royal Astronomical Society},
  year = {2003},
  volume = {339},
  number = {3},
  pages = {793–809},
  doi = {10.1046/j.1365-8711.2003.06224.x}
}

@ARTICLE{2016ApJ...821...54S,
       author = {{Sonbas}, E. and {Rangelov}, B. and {Kargaltsev}, O. and {Dhuga}, K.~S. and {Hare}, J. and {Volkov}, I.},
        title = "{X-Ray Sources in the Dwarf Spheroidal Galaxy Draco}",
      journal = {\apj},
         year = 2016,
        month = apr,
       volume = {821},
       number = {1},
          eid = {54},
        pages = {54},
          doi = {10.3847/0004-637X/821/1/54},
archivePrefix = {arXiv},
       eprint = {1505.00216},
 primaryClass = {astro-ph.GA},
       adsurl = {https://ui.adsabs.harvard.edu/abs/2016ApJ...821...54S}
}

@ARTICLE{2016ApJ...816...52H,
       author = {{Hare}, Jeremy and {Rangelov}, Blagoy and {Sonbas}, Eda and {Kargaltsev}, Oleg and {Volkov}, Igor},
        title = "{Multi-wavelength Study of HESS J1741-302}",
      journal = {\apj},
         year = 2016,
        month = jan,
       volume = {816},
       number = {2},
          eid = {52},
        pages = {52},
          doi = {10.3847/0004-637X/816/2/52},
archivePrefix = {arXiv},
       eprint = {1506.05736},
 primaryClass = {astro-ph.HE},
       adsurl = {https://ui.adsabs.harvard.edu/abs/2016ApJ...816...52H}
}
\bibliographystyle{aasjournalv7}



\end{document}